\documentclass[11pt]{article}

\usepackage{acl}

\usepackage{times}
\usepackage{latexsym}

\usepackage[T1]{fontenc}

\usepackage[utf8]{inputenc}

\usepackage{microtype}

\usepackage{inconsolata}

\usepackage{graphicx}

\usepackage{multirow}

\usepackage{algorithm}
\usepackage{algorithmic}
\usepackage{adjustbox}
\usepackage{array}
\newcolumntype{x}[1]{>{\centering\let\newline\\\arraybackslash\hspace{0pt}}p{#1}}
\usepackage{booktabs}
\usepackage{comment}
\usepackage[inline]{enumitem}
\usepackage{etoolbox}
\usepackage[utf8]{inputenc}
\usepackage{makecell}
\usepackage{mathtools}
\usepackage{multirow}
\usepackage{nicefrac}
\usepackage{soul}
\usepackage{subcaption}
\usepackage{tabularx}
\usepackage[table]{xcolor}
\usepackage[capitalise,noabbrev]{cleveref}
\usepackage{newfloat}
\usepackage{listings}
\DeclareCaptionStyle{ruled}{labelfont=normalfont,labelsep=colon,strut=off} 
\floatstyle{ruled}
\newfloat{listing}{tb}{lst}{}
\floatname{listing}{Listing}

\title{C-QUERI: Congressional Questions, Exchanges, and Responses in Institutions Dataset}

\author{
 \textbf{Manjari Rudra\textsuperscript{1}},
 \textbf{Daniel Magleby\textsuperscript{2}},
 \textbf{Sujoy Sikdar\textsuperscript{1}},
\\
 \textsuperscript{1}School of Computing, Binghamton University, Binghamton, New York
\\
 \textsuperscript{2}Department of Political Science, Binghamton University, Binghamton, New York
\\
}

\begin{document}
\maketitle
\begin{abstract}
Questions in political hearings serve strategic purposes beyond information gathering, including advancing partisan narratives and shaping public perceptions, yet remain understudied at scale. Congressional hearings provide an especially rich and tractable site for studying political questioning: Interactions are structured by formal rules, witnesses are obliged to respond, and members with different political affiliations are guaranteed opportunities to ask questions.

We develop a pipeline to extract question-answer pairs from unstructured hearing transcripts and construct a novel dataset of committee hearings from the 108th--117th Congress. Our analysis reveals systematic differences in questioning strategies across parties, showing that questioner party affiliation can be predicted from questions alone. Our dataset and methods advance the study of congressional politics and provide a framework for analyzing question-answering across interview-like settings.
\end{abstract}


\section{Introduction}

Congressional \emph{hearings} are among the most visible and consequential activities of the U.S. Congress \citep{Ban_Park_You_2024, BAN_HILL_2025}. In these forums, elected representatives question witnesses including cabinet officials, agency heads, policy experts, and industry leaders. Hearings shape the legislative agenda, provide oversight of executive agencies, and serve as a stage for partisan conflict and coalition-building. This offers the public a rare window into the complex, and often contested, processes of deliberation, lawmaking, investigation, and accountability at the highest levels of government. Yet, despite their centrality to democratic accountability, the language of question-answering in hearings has been difficult to study at scale. While transcripts are public, they rarely appear in a form that supports utterance and exchange level analysis of who asked what, how questions were framed, and how witnesses responded.

This paper introduces a new dataset of question-and-answer exchanges from congressional hearings, together with a computational pipeline for constructing it. We segment hearing transcripts into utterances, identify question-answer exchanges using fine-tuned transformer-based classifiers, and pair questions with the responses they elicit. We validate key pipeline steps using human annotation and provide a rich set of linguistic features and metadata to support downstream analysis and natural language processing tasks.

As partisan divisions have deepened, hearings have become sites of political performance as much as information gathering \citep{LevitskyZiblatt2018, Klein2020}. Prior research on congressional partisanship has focused on floor votes and roll-call data, which capture only issues that reach the floor and obscure upstream processes in which committees decide what gets debated \citep{mayhew1974congress, coxmccubbins1993, coxmccubbins2005}. Hearings offer a complementary window: members with different affiliations are guaranteed opportunities to ask questions, witnesses are obliged to respond, and questioning alternates across majority and minority members, enabling direct comparisons across legislators, parties, and time.

We hypothesize that political agendas and strategic goals are reflected in the linguistic structures, tone, and style of congressional questioning. Our central research question is: \textbf{Do party affiliation and majority/minority status affect the form and content of questions and the answers they elicit in congressional hearings?}

\subsection{Our Contributions}

Our contributions are three-fold. First, we introduce a novel dataset constructed from transcripts of 16,130 committee hearings in the U.S. House and Senate spanning the 108th to 117th Congresses, from which we extract over 3 million utterances. The dataset identifies question and answer utterances, includes human annotations to validate classification, and provides linguistic features for each exchange, opening new avenues for studying congressional behavior, polarization, and question-answer dynamics at scale.

Second, we develop a pipeline for processing unstructured hearing transcripts into analyzable question-answer units. The pipeline combines BERT-based and heuristic named entity recognition for speaker identification, fine-tuned BERT classifiers for question-answer labeling, and a comprehensive linguistic feature extraction step. Beyond congressional hearings, this framework generalizes to other structured interview-like settings including media interviews and judicial proceedings, enabling comparative studies of political questioning across institutions.

Third, we provide an empirical assessment of partisan signals in hearing dialogue using BERT-based classifiers, classical machine learning on human-engineered linguistic features, and zero-shot prompting with large language models. Across all hearings, we predict questioner party affiliation with 59\% accuracy which is 8\% above the majority-class baseline, showing that partisanship permeates not only voting but also deliberative and oversight functions of Congress. Notably, aggregate accuracy masks dramatic variation: performance ranges from near-random in some committees to over 70\% in others. We show that the ability to distinguish party affiliation varies systematically with committee, hearing type, and whether government is unified or divided. These findings speak to longstanding debates about whether congressional committees primarily serve informational or partisan ends \citep{krehbiel1991, coxmccubbins2005}.

\section{Related Work}

\noindent{\bf Legislative and Political Discourse Datasets.} Several datasets provide access to transcripts of floor speeches and debates in the U.S. Congress~\citep{gentzkow2018congressional}, European democracies~\citep{rauh2020parlspeech, Norwegian_fiva2025}, and local council meetings~\citep{ConstituentsDataset_Hoang2025}. \citet{delanobulk} archive congressional data including committee hearings and reports. The Congressional Committees Hearing dataset (CoCoHD) is the most similar dataset to ours, consisting of transcripts from the 105th to the 118th sessions~\citep{CoCoHD_hiray2024}. Our dataset differs from CoCoHD in three ways: we segment transcripts into individual utterances, identify question and answer utterances within each hearing, and pair questions with their corresponding answers to enable downstream question-answer analysis. We also provide utterance-level human annotations and a comprehensive set of linguistic features. \citet{GermanPolInterview_Birkenmaier2025} extract transcripts of political interviews from news media, and \citet{SocialTestimony_maher2020} analyze testimonies of social scientist witnesses from 1946 to 2016. None of these datasets identify question and answer utterances for large-scale computational analysis.

\noindent{\bf Interview and Dialogue Datasets.} The MediaSum~\citep{zhu2021mediasum} and Interview~\citep{majumder-etal-2020-interview} datasets consist of interview dialogues from mainstream media, constructed primarily for summarization tasks. While these share the interview-like structure of congressional hearings, they do not distinguish question from answer utterances or provide speaker metadata linking participants to institutional roles. \citet{thomas2024never} introduce a dataset of Q\&A pairs from political interviews with U.S. presidents, annotated with a taxonomy of response clarity and evasion techniques. \citet{PoliticalPolarizationUSC_Fisher2013} examine statements by committee members from climate-related hearings using discourse network analysis. Our dataset complements these by enabling large-scale, utterance-level analysis of question-answer exchanges across a broad range of committees and sessions.

\noindent{\bf Question Analysis in NLP.} The NLP question-answering literature focuses predominantly on information-seeking questions, evaluated on extracting correct responses from text or knowledge bases~\citep{zhang2023survey, biancofiore2024interactive, yu2024natural}, conversational question answering~\citep{zaib2022conversational}, and natural language understanding~\citep{kovcisky2018narrativeqa}. Most closely related to our work, \citet{zhang_2017_RhetoricalRoles} introduce an unsupervised framework for identifying rhetorical question types in UK parliamentary discourse, revealing systematic differences between government and opposition questioners. \citet{ferracane2021did} study conversation acts and intents in 20 U.S. congressional hearings, examining whether witnesses answer the questions posed to them and how question form shapes responses. Our dataset extends this line of work to a corpus three orders of magnitude larger, pairing questions with answers across all major committees and congressional sessions, and providing the infrastructure for studying partisan signals in questioning strategies at scale. More broadly, NLP research has examined rhetorical question identification~\citep{bhattasali2015automatic} and dialogue act classification in naturalistic conversational settings~\citep{stolcke2000dialogue, raheja2019dialogue}. Questions in congressional hearings, however, are embedded in a distinct institutional context where they are frequently rhetorical, adversarial, or performative. A member asking ``Isn't it true that your agency failed to act?'' is making an accusation in interrogative form, not seeking information. Our dataset provides infrastructure for studying such questions at scale, complementing existing question-answering paradigms.

\noindent{\bf Computational Analysis of Political Text.} A substantial body of NLP work has analyzed political discourse, including ideology detection from text~\citep{thomas2006get, iyyer2014political}, framing and agenda-setting~\citep{card2015media}, and partisan language in floor speeches~\citep{gentzkow2019, lauderdale2016}. Text-as-data approaches have tracked polarization in congressional speech and linked linguistic features to roll-call voting patterns. Our work extends this line of research from floor speeches to committee hearings, where the structured question-answer format enables finer-grained analysis of how partisanship is expressed in dialogue rather than monologue.

\noindent{\bf Congressional Hearings in Political Science.} Congressional hearings are a central but understudied arena of legislative politics. Prior research on oversight has relied on aggregate measures such as hearing counts~\citep{kriner2008, kriner2016}, and classic work emphasizes that congressional monitoring of the executive is difficult to observe and episodic at best~\citep{mccubbins1984, abercbach1990}. Recent work has begun to examine how hearings shape bureaucratic behavior and how partisan incentives determine from whom Congress seeks information~\citep{Ban_Park_You_2024, BAN_HILL_2025}. Informational theories of Congress~\citep{krehbiel1991, gailmard2013} and work on descriptive representation~\citep{mansbridge1999, swers2002, swers2020, kanthak2013, kathlene1994, minta2011} have lacked direct measures of the exchanges they theorize, and concerns have been raised about the underrepresentation of women among congressional witnesses~\citep{Underrepresentation_Coil2024}. Studies of polarization have focused on floor votes and speeches~\citep{mccarty2006, lee2016}, leaving oversight hearings largely unexamined, while theories of deliberative democracy~\citep{gutmann1996, bessette1994, mansbridge2012} lack direct empirical grounding in legislative practice. By structuring hearings into discrete question-answer exchanges, our dataset makes legislative dialogue observable at a fine-grained level, enabling systematic tests of when hearings generate genuine policy deliberation and when they serve partisan signaling functions.

\section{Constructing the C-QUERI Dataset}\label{sec:congressional}
\begin{figure}[h]
    \centering
    \includegraphics[width=1\linewidth]{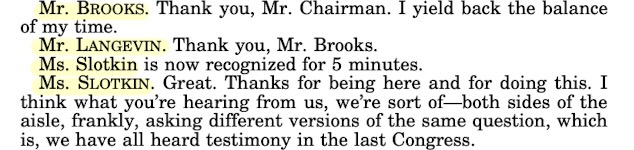}
    \caption{Excerpt of a congressional hearing transcript\label{fig:trans}}
\end{figure}

Public Congressional hearings are made available in digital format at \url{https://www.govinfo.gov/app/collection/chrg} between 2 months to 2 years after the hearing is held, and as Fig.\ref{fig:trans} illustrates, are published as a dialogue style transcript in plain text. Our dataset\footnote{\url{https://doi.org/10.5281/zenodo.15470512}} is constructed from all available transcripts as of April 2022, covering 20 years spanning the 108th to the 117th congressional sessions with transcripts from 16,130 hearings by 76 committees involving 1,774 committee members and 69,149 witnesses in total. 

Fig.~\ref{fig:hearingshm} and~\ref{fig:utteranceshm} in Appendix~\ref{app:Supp_Dataset_Statistics} illustrates the distribution of hearings, witnesses, utterances and word-counts across congressional sessions and committees, revealing trends that track major domestic and global events.

We note that some committees have been renamed over time although their roles and responsibilities typically have not changed. For example, the House Committee on Foreign Affairs operated as the Committee on International Relations from 1975 to 1978 and 1995 to 2007. To avoid confusion, we maintain the name under which the committee operated at the time the hearing was conducted and only consider activity across committees when appropriate for further analysis.

\noindent{\bf Transcripts.} 
Congressional hearings follow a common format that is reflected in the transcripts which are published as a single, unstructured text document. A hearing typically starts with a roll call of members present, followed by an address by the committee chair, and statements from committee members and witnesses. This is followed by exchanges where committee members ask questions which are answered by the witnesses. Some hearings have more structure with fixed time slots allocated to each member. Metadata published in the Extensible Markup Language (XML) identifies the hearing, committee(s), members in attendance, and witnesses present at the hearing along with their affiliations. We observe that there can be significant delays in the publication of the official transcripts and the metadata may sometimes fail to identify committee members and witnesses in attendance.
Curating our dataset consisting of individual utterances together with linguistic features and identifying questions and answers from transcripts in plain text involves the following tasks.

\subsection{Task 1: Identify Utterances}
We identify individual utterances in a transcript using the following pipeline.

\noindent{\bf Preprocessing.} We identify the beginning and end of proceedings using rules developed through careful observation of the structure of a transcript, allowing us to remove parts of the transcript that convey redundant metadata. Detailed documentation of these rules accompanies the dataset.

\noindent{\bf Extracting utterances.} We use the word ``utterance'' to mean the consecutive words spoken by a single member or witness. As Figure \ref{fig:trans} illustrates, the proceedings are transcribed so that the name of the speaker (member or witness) precedes their utterance, a structure that we exploit using named entity recognition. A key challenge we encounter is that standards for formatting these markers vary by committee and the time at which the transcript was produced. For example, honorifics, full name or last name alone, case and capitalization, and speaker's affiliation are used inconsistently across the raw transcripts.

\noindent{\bf Named Entity Recognition.}
We develop a hybrid Named Entity Recognition (NER) method to identify these markers by combining large language model (LLM) based and heuristic techniques. We use a BERT-based NER following \citet{devlin2018BERT} to identify and classify entities, such as names of people, organizations, locations, dates, and other specific items within a text. To augment this, we develop heuristic NER rules by systematic examination of transcript and metadata formatting patterns which we implement using regular expressions. We tag named entities appearing at the beginning of utterances as either a member of the committee or a witness using data obtained from the metadata associated with the hearing, and through the Propublica Congress API (\url{https://projects.propublica.org/api-docs/congress-api/members/}) which provides details about members their roles and affiliations.   

Code to reproduce results in this submission can be found at  \url{https://anonymous.4open.science/r/CongressHearings-27B6/README.md} and data can be found at \url{https://tinyurl.com/3yudwner}. 

\subsection{Task 2: Identify Questions and Answers}\label{subsec:qna}

Next, we identify question and answer utterances and focus particularly on the question and answer sessions that follow the statements by the witnesses. We take a supervised transfer learning approach to fine-tune a variant of the BERT model obtained from Huggingface \footnote{\url{https://huggingface.co/google-bert/bert-base-cased}} coupled with a dropout layer and a linear layer with ReLU activation in order to classify the utterances as questions and answers (Figure \ref{fig:BERT} in Appendix \ref{app:LLM_Config}). 

Since transcripts of Congressional committee hearings are not annotated, we construct a training dataset from the Reddit Ask Me Anything (r/AMA) community consisting of 60756 Questions and 60756 Answers, and the U.K. Parliamentary dataset~\cite{zhang_2017_RhetoricalRoles} with 2344 questions and 2344 answers. The r/AMA community lets individuals field questions from Reddit users, with top-ranked comments serving as questions and the poster's responses as answers; we construct our training set from the highest-ranked exchanges. The U.K. Parliamentary dataset consists of written questions by members of the U.K. Parliament directed at cabinet ministers who are typically required to respond within a set timeframe. Although these training sources differ in register from congressional hearings, our classifier achieves 87\% accuracy on held-out congressional utterances, validating the transfer. Details are in Appendix~\ref{app:QnA}.

    \subsection{Dataset Curation Results}
        \subsubsection{Task 1: Identify Utterances.} \label{sec:UtteranceRes}
        Our dataset consists of 3,319,386 utterances. We evaluate the efficacy of our utterance identification using an annotated evaluation dataset constructed by first sampling 50 hearings from each congressional session uniformly at random. From each of these hearings, we sample 10 utterances from the output of our utterance identification pipeline, for a total of 500 utterances from each congressional session. We then manually verify whether these utterances were correctly identified.

        Overall, 93.96\% of the utterances identified by our pipeline were correct in our manual verification. Incorrectly split utterances occur due to one of two kinds of error; utterances that are {\em clubbed} together or a single utterance that is {\em broken} and split in two or more utterances. Table~\ref{tab:Task1_Result} summarizes our findings for the task of identifying utterances. 

        \begin{table}
            \centering
            \footnotesize
            \setlength\tabcolsep{1pt}
            
            \begin{tabular}{|c|c|c|c|}
                \hline
                \textbf{Session} & \textbf{\# Incorrect} & \textbf{\# Clubbed} & \textbf{\# Broken} \\
                \hline
                108 & 32 & 12 & 20 \\
                109 & 16 & 8 & 8 \\
                110 & 20 & 15 & 5 \\
                111 & 12 & 8 & 4 \\
                112 & 4 & 2 & 2 \\
                113 & 39 & 36 & 3 \\
                114 & 27 & 18 & 9 \\
                115 & 23 & 18 & 5 \\
                116 & 52 & 40 & 12 \\
                117 & 77 & 69 & 8 \\
                \hline \hline
                \textbf{Total} & \textbf{302} & \textbf{226} & \textbf{76} \\
                \hline
            \end{tabular}
            \caption{Performance on utterance identification (Task 1). We sample 50 hearings per session, from each of which we sample 10 utterances and manually verify them. Over 10 sessions and 5,000 verified utterances, 302 were incorrectly identified, of which 226 were clubbed and 76 were broken.}
            \label{tab:Task1_Result}
        \end{table}
        
        \subsubsection{Task 2: Identify Questions and Answers.} \label{sec:QuestionAnswer}
        To evaluate our question and answer annotations, we sampled 800 utterances uniformly at random from among all utterances that occurred between the 114th through to the 117th congressional sessions. These 800 utterances were then hand-labeled, yielding 379 questions and 421 answers. We achieved an accuracy of 87.14\% at the question-answer labeling task using our BERT-based classifier. Table \ref{tab:QnAResult} shows the confusion matrix for this task. Overall, 291 and 380 out of the 379 and 421 question and answer utterances respectively were classified correctly.

        \begin{table}
            \centering
            \footnotesize
            \setlength\tabcolsep{3pt}
            \begin{tabular}{ll|cccc|c|}
                \hline
                \multicolumn{2}{|r|}{\bf Session} & {\bf 114}     & {\bf 115}     & {\bf 116}     & {\bf 117}    & {\bf 114--117} \\ 
                \hline \hline
                \multicolumn{1}{|l|}{\multirow{2}{*}{\bf Questions}} & True  & 20      & 105     & 101     & 65     & 291 \\ 
                \cline{2-7}
                \multicolumn{1}{|l|}{}                          & False & 6       & 25      & 29      & 28     & 88 \\ 
                \hline
                \multicolumn{1}{|l|}{\multirow{2}{*}{\bf Answer}}   & True  & 26      & 127     & 134     & 93     & 380 \\ 
                \cline{2-7}
                \multicolumn{1}{|l|}{}                          & False & 0       & 3       & 3       & 5      & 11 \\ 
                \hline \hline
                \multicolumn{2}{|l|}{\bf Accuracy}                          & {\bf 0.88} & {\bf 0.89} & {\bf 0.87} & {\bf 0.83} & {\bf 0.87} \\ \hline
            \end{tabular}
            \caption{Performance on identifying question and answer utterances (Task 2), against 800 human labeled utterances. For each session indicated by a column, a row entry provides the number of utterances either correctly (True) or incorrectly (False) identified as a question or answer. Over sessions 114-117, 291 out of 379 utterances identified as questions by our classifier were truly question utterances.}
            \label{tab:QnAResult}
        \end{table}

\section{Methodology for Identifying Party Affiliation and Standing} \label{sec:PartyPrediction}
    The task is to predict the questioner's {\em party affiliation}, either Democrat (D), Republican (R), or Independent (I), or {\em party standing}, i.e., whether their party holds a majority (M) or minority (m) in their respective chamber, from the text of a question utterance, an answer utterance, or both. Each data instance is a single utterance labeled with the speaker's affiliation or standing. We take a three-pronged approach, with each method offering distinct advantages: classical machine learning provides interpretable, human-readable feature importance; fine-tuned BERT captures deep contextual semantics; and zero-shot LLM prompting establishes whether partisan signals are detectable without task-specific training.
    
    \noindent{\bf Approach 1: Classical ML with linguistic features.} We train random forest classifiers on a comprehensive suite of human-engineered features from the News Landscape (NELA) toolkit~\citep{horne_2020_NELA}, including \textbf{Style} (POS tags, punctuation usage), \textbf{Complexity} (lexical diversity, reading difficulty, sentence length), \textbf{Affect} (sentiment), and \textbf{Bias} (hedge words, factives, assertatives). Random forests achieved the best performance among classical methods including SVMs and logistic regression variants. We use an 80/20 train/test split with 5-fold cross-validation for hyperparameter tuning (Appendix.~\ref{app:Gridsearch}), and use feature importance scores to identify the most significant features and study class differences using Kolmogorov-Smirnov tests (Appendix~\ref{app:c}).
        
    \noindent{\bf Approach 2: Fine-tuned BERT Classifier.} We fine-tune a BERT-based classifier~\citep{devlin2018BERT} consisting of the BERT encoder followed by a dropout layer and a linear layer with LeakyReLU activation using the cross-entropy loss function and the Adam optimizer. We use a 20/5/75 train/validation/test split with 5-fold cross-validation to select hyperparameters including batch size, learning rate, and number of epochs. We train separate classifiers for each combination of congressional session and committee to capture context-specific partisan signals. Architecture details are in Appendix~\ref{app:LLM_Config}.

    \noindent{\bf Approach 3: Zero-shot LLM Prompting.} We evaluate the ability of a wide range of popular pre-trained LLMs to distinguish party affiliation and standing using zero-shot prompting techniques. Details are presented in Appendix~\ref{app:LLM_Config}.
    
\section{Results} \label{sec:PartyPrediction_result}
    We begin by testing whether the party affiliation and standing of members of Congress can be distinguished using linguistic features and semantic content of the questions they ask and the answers they receive over the entire dataset of all hearings. Our results for predicting party affiliation and standing are summarized in Table~\ref{tab:Party_Results}.

    \begin{table}[h]
        \footnotesize
        \centering
        \begin{tabular}{|l|c|c|c|c|}
            \cline{2-5}
            \multicolumn{1}{c|}{} & \multicolumn{2}{c|}{\textbf{Affiliation}} & \multicolumn{2}{c|}{\textbf{Standing}}\\\hline
            \textbf{Model} & \textbf{Q} & \textbf{A} & \textbf{Q} & \textbf{A} \\ \hline \hline
            \multicolumn{1}{|p{2.5cm}|}{NELA}         & \textbf{53.75} & \textbf{52.22} & \textbf{61.05} & 60.97 \\\hline
            \multicolumn{1}{|p{2.5cm}|}{NELA Baseline}& 51.52 & 51.53 & 60.97 & 60.97 \\\hline \hline
            \multicolumn{1}{|p{2.5cm}|}{BERT w/o speaker}  & \textbf{59.04} & \textbf{54.70} & 58.90& 57.39 \\\hline
            \multicolumn{1}{|p{2.5cm}|}{BERT Baseline}& 51.66 & 51.62& 61.45 & 61.45 \\\hline
        \end{tabular}
        \caption{Results for identifying the questioner's party affiliation(Republican vs Democrat) and standing (majority vs minority) when the input is a question (Q) or the answer (A) in response to a question. Speaker names were removed from utterances unless otherwise specified. Base represents the fraction of test examples with the majority class.}
        \label{tab:Party_Results}
    \end{table}

    \noindent{\bf Question-based prediction.} After removing names of committee members and witnesses, which appear at the beginning of utterances in the transcript, we fine-tune the BERT model. We find that party affiliation can be predicted with an accuracy of 59\%, i.e., 8\% higher than the baseline of predicting the majority class in the test set. This indicates subtle semantic differences between the content of questions posed by members with different party affiliations. In contrast, zero-shot prompting with LLMs on a random sample of 10,000 question-answer pairs fails to exceed baseline performance, as we show in Table~\ref{tab:Party_Results_LLM} in Appendix~\ref{app:SupResults} indicating that these partisan linguistic patterns may be too nuanced for detection without fine-tuning. For party standing, we are unable to beat the baseline using a similar approach. 

    Notably, when speaker names are retained in the question text, party affiliation prediction accuracy jumps to 92\%. This suggests that embeddings of committee members' names capture systematic differences in how politicians from different parties are referenced in text corpora, reflecting partisan patterns in media coverage, political discourse, and public communication, but are unable to resolve linguistic or semantic differences when trained on all committee hearings. This underscores the importance of name removal in isolating genuine differences in questioning style. Therefore, we only report results on input utterances with speaker names removed, unless we mention otherwise.

    \begin{table*}[htp]
        \centering
        \footnotesize
        \begin{tabular}{|c|c|c|c|l|}
            \hline
            \textbf{BERT} & \textbf{BERT Baseline} & \textbf{NELA} & \textbf{NELA Baseline} & \textbf{Committee} \\\hline
             \textbf{64.48} & 54.75(R) & \textbf{58.73} & 54.71(R) & (H) Energy and Commerce\\ \hline
             \textbf{65.39} & 56.40(R) & 57.85 & 56.38(R) & (H) Financial Services\\ \hline 
             \textbf{62.54} & 55.19(D) & \textbf{57.10} & 54.87(D) & (H+S) Judiciary\\ \hline
             \textbf{72.48} & 70.21(R) & 71.07 & 70.42(R) & (H) Oversight and Govt. Reform\\ \hline
             64.69 & 66.47(R) & \textbf{68.76} & 66.91(R) & (H) Ways and Means\\ \hline
             \textbf{67.13} & 63.99(D) & 63.66 & 63.69(D) & (S) Homeland Security and Govt. Affairs\\ \hline
             60.95 & 60.54(D) & 62.41 & 61.47(D)& (S) Comm, Sci, and Transportation\\ \hline
             \textbf{59.27} & 51.20(D) & \textbf{54.81} & 51.35(D) & (H) Armed Services \\ \hline
             \textbf{61.04} & 50.32(D) & \textbf{54.88} & 50.26(D) & (H+S) Veterans' Affairs \\ \hline
             \textbf{60.00} & 50.80(R) & \textbf{57.72} & 51.79(R) & (H) Foreign Affairs \\ \hline
        \end{tabular}
        \caption{Results for identifying the questioner's party affiliation (Republican vs Democrat) on question utterances across the 10 largest committees by number of committee hearings using BERT and Random Forest Classifier. An (R) or (D) indicates that Republican or Democrat speakers have more utterances, and H = House, S = Senate.}
        \label{tab:ParQBERT}
    \end{table*}

    \noindent{\bf Answer-based prediction.} Identifying the party affiliation or standing of the questioner based on answer utterances appears to be more challenging as we achieve accuracy that is only 1\% and 0\% above their respective baselines.

    \noindent{\bf Linguistic features.} Classical linguistic features from the NELA toolkit alone are insufficient to predict either party affiliation or standing (Table~\ref{tab:Party_Results}). 
    
    We further explore party differences in linguistic style by applying Kolmogorov–Smirnov tests across NELA features, as we discuss in Appendix~\ref{app:c} in greater detail. Our feature-level analysis highlights statistically significant linguistic differences between parties. Republicans rely more heavily on assertive and hedge words, whereas Democrats more frequently employ positive sentiment and implicative phrasing. These distinctions suggest that partisan communication styles are not only detectable computationally but also reflect broader rhetorical strategies consistent with each party’s issue framing and public appeals.

    \subsection{Differences Occur in Certain Committees and Hearing Types}

    A natural follow-up question is whether party affiliation or standing is more distinguishable under specific contexts. We examine this question along three dimensions: 
    \begin{enumerate*}[label=(\arabic*),leftmargin=*,topsep=0pt,itemsep=0pt]
        \item {\em individual committees} and Congressional sessions,
        \item {\em types of hearings} and legislative functions, and
        \item across times of {\em unified} or {\em divided} government.
    \end{enumerate*} Here, government is unified during a given Congressional session if the same party controlled the presidency as well as both the House of Representatives and the Senate.

    \noindent{\bf Differences across Committees.}
    Table~\ref{tab:ParQBERT} (and Figure~\ref{fig:Committee_Accuracy} in Appendix~\ref{app:SupResults}) reveals substantial variation in how well classifiers predict party affiliation across committees. We focus on the improvement over the majority-class baseline, indicating how much partisan signal exists beyond mere class imbalance, by which committees roughly fall into three tiers.

    The strongest partisan signals ($\ge$+8.0) emerge in Veterans' Affairs, Energy and Commerce, Foreign Affairs, Financial Services, and Armed Services. A plausible explanation is that both parties often share broad headline goals (e.g., supporting veterans or national security) but differ on how they frame issues, have different priorities, invoke different values, and attribute problems to different causes, creating linguistically distinct questioning even when substantive positions overlap. This is consistent with theories of issue ownership \citep{egan2013}, issue frame ownership \citep{arbour2014issue}, and party reputation and branding that highlight collective incentives for coordinated messaging \citep{coxmccubbins1993,coxmccubbins2005}.
    
    We observe moderate partisanship ($<$8.0 and $\ge$3.0) in Judiciary, and Homeland Security and Governmental Affairs, which handle issues where partisan positions are well-established, but policy complexity and legal constraints may impose some convergence in questioning style.

    The weakest partisan signals are observed in Oversight and Government Reform, Ways and Means, and Commerce, Science, and Transportation. This is initially surprising, because Oversight and Ways and Means are among the most visibly partisan arenas in Congress. A possible explanation is that in these high-stakes settings, both parties adopt similar adversarial or technically focused questioning strategies. For example, Oversight hearings often feature aggressive questioning and demands for accountability regardless of party, and the partisan difference may lie in whom members target rather than how they question. Ways and Means deals with tax policy where questioning may be constrained by well-established institutional norms~\citep{fenno1973}.

    Our findings point to the linguistic imprint of partisanship being strongest not where conflict is most intense, but where parties frame the issues through different rhetorical strategies. This complicates simple mappings between committee salience and partisan distinctiveness. High-profile committees often viewed as central to party reputations do not necessarily produce the most distinguishable questioning styles. Instead, partisan linguistic signals may reflect framing differences more than conflict intensity, a pattern consistent with~\citet{lee2009}'s argument that congressional partisanship is often driven by electoral incentives for differentiation rather than substantive policy disagreement.
    
    The ability of linguistic features 
    to predict party affiliation in questions for certain combinations of sessions and committees (Table~\ref{tab:ParQNELA}, Appendix~\ref{app:SupResults}) suggests that the partisan imprint on questioning may fluctuate with broader institutional and electoral contexts, echoing findings in the literature on conditional party government \citep{aldrichrhode2000} and partisan messaging \citep{lee2009}.

       \begin{table}[h]
        \centering
        \footnotesize
        \setlength\tabcolsep{3pt}
        \begin{tabular}{|l|l|cc|}
            \hline
            \textbf{Committee on}  & \textbf{Govt.} & {\bf BERT}&{\bf Baseline} \\ 
            \hline
            \multicolumn{1}{|l|}{\multirow{2}{*}{(H)\textbf{Energy and Commerce}}}  
                                    & Unified   & \textbf{60.33} & 52.22(D) \\ 
            \cline{2-4}
            \multicolumn{1}{|l|}{}  & Divided   & \textbf{65.53} & 56.57(R) \\
            \hline
            \multicolumn{1}{|l|}{\multirow{2}{*}{(H) \textbf{Financial Services}}}    
                                    & Unified   & \textbf{61.20} & 54.05(R) \\
            \cline{2-4}
            \multicolumn{1}{|l|}{}  & Divided   & \textbf{64.88} & 59.15(R) \\ 
            \hline
            \multicolumn{1}{|l|}{\multirow{2}{*}{(H+S) \textbf{the Judiciary}}}
                                    & Unified   & \textbf{58.93} & 56.23(D) \\
            \cline{2-4}
            \multicolumn{1}{|l|}{}  & Divided   & \textbf{60.66} & 55.72(D) \\
            \hline
            \multicolumn{1}{|l|}{\multirow{2}{*}{(H) \textbf{Oversight and Gov Reform}}}
                                    & Unified   & 66.50 & 68.17(R) \\
            \cline{2-4}
            \multicolumn{1}{|l|}{}  & Divided   & 71.20 & 69.34(R) \\
            \hline
            \multicolumn{1}{|l|}{\multirow{2}{*}{(S) \textbf{Homel Sec and Gov Affairs}}}   
                                    & Unified   & \textbf{63.16} & 60.15(D) \\
            \cline{2-4}
            \multicolumn{1}{|l|}{}  & Divided   & 67.08 & 66.72(D) \\
            \hline
            \multicolumn{1}{|l|}{\multirow{2}{*}{(S) \textbf{Comm, Sci, and Transports}}}  
                                    & Unified   & 59.64 & 58.28(D) \\
            \cline{2-4}
            \multicolumn{1}{|l|}{}  & Divided   & 52.42 & 65.24(D) \\
            \hline
            \multicolumn{1}{|l|}{\multirow{2}{*}{(H) \textbf{Armed Services}}}   
                                    & Unified   & \textbf{59.66} & 54.96(D) \\
            \cline{2-4}
            \multicolumn{1}{|l|}{}  & Divided   & 50.60 & 50.61(R)\\
            \hline
            \multicolumn{1}{|l|}{\multirow{2}{*}{(H+S) \textbf{Veterans' Affairs}}}   
                                    & Unified   & \textbf{55.06} & 52.20(D) \\
            \cline{2-4}
            \multicolumn{1}{|l|}{}  & Divided   & \textbf{60.34} & 52.74(R) \\
            \hline
            \multicolumn{1}{|l|}{\multirow{2}{*}{(H) \textbf{Foreign Affairs}}}   
                                    & Unified   & 48.31 & 58.79(D) \\
            \cline{2-4}
            \multicolumn{1}{|l|}{}  & Divided   & \textbf{60.87} & 57.31(R) \\
            \hline
        \end{tabular}
        \caption{Predictions of party affiliations from question utterances across committees, comparing unified and divided Congress settings. An annotation of (R) or (D) indicates that Republican or Democrat speakers have the most utterances.}
        \label{tab:UniDivGovPartyResultsC_BERT}
    \end{table}
    
    \noindent{\bf Differences across Government Type.} The findings reported in Table~\ref{tab:UniDivGovPartyResultsC_BERT} and Table~\ref{tab:UniDivGovPartyResultsC_LLM} in Appendix~\ref{app:SupResults}, obtained by fine-tuning a BERT-based classifier and zero-shot prompting respectively, provide insights that connect ongoing debates in congressional scholarship. First, partisan identity shows the strongest classifier improvement over baseline in committees such as Energy and Commerce and Financial Services, where regulatory debates often divide along party lines, consistent with theories of partisan agenda control \citep{coxmccubbins2005,lee2009}. Judiciary also shows consistent improvement across both government configurations. Second, classifiers perform inconsistently in committees associated with distributive politics or bipartisan traditions: Armed Services and Veterans' Affairs show stronger signal under unified and divided governments, respectively. Third, while many committees show stronger partisan signals under divided government, consistent with findings on the dynamics of contested institutional control \citep{mayhew2005divided}, there are notable exceptions. This suggests that the relationship between institutional context and partisan language is committee and context specific rather than uniform.


\begin{table}[h]
        \centering
        \footnotesize
        \setlength\tabcolsep{3pt}
        \begin{tabular}{|l|l|cc|}
            \hline
            \textbf{\bf Hearing Type} & \textbf{Government}  & {\bf BERT}&{\bf Baseline} \\ 
            \hline 
            \multicolumn{1}{|l|}{\multirow{3}{*}{\bf General}} & All & \textbf{58.42} & 51.51(R) \\\cline{2-4}
                                                            & Unified         & \textbf{51.72} & 50.46(D)\\\cline{2-4}  
            
            \multicolumn{1}{|l|}{}                        & Divided         & \textbf{59.57} & 52.14(R)\\      
            \hline
            \multicolumn{1}{|l|}{\multirow{2}{*}{\bf Field}} & All & \textbf{57.21} & 54.90(R) \\\cline{2-4}
                                                                & Unified         & 53.34 & 53.30 (R)\\\cline{2-4}    
            \multicolumn{1}{|l|}{}                              & Divided         & \textbf{55.90} & 50.47(R)\\\hline
            \multicolumn{1}{|l|}{\multirow{2}{*}{\bf Oversight}} & All & \textbf{61.36} & 58.15(R) \\\cline{2-4}
                                                                & Unified         & \textbf{59.36} & 56.98(R)\\\cline{2-4} 
            \multicolumn{1}{|l|}{}                              & Divided         & 67.25 & 70.45(D)\\ \hline
            \multicolumn{1}{|l|}{\multirow{2}{*}{\bf Authorization}} & All & 51.9 & 51.81(R) \\ \cline{2-4}
                                                                & Unified     & \textbf{53.65} & 51.16(D)\\\cline{2-4} 
            \multicolumn{1}{|l|}{}                              & Divided         & \textbf{52.07} & 50.15(R)\\\hline
            \multicolumn{1}{|l|}{\multirow{2}{*}{\bf Nomination}} & All & 58.90 & 58.16(D) \\\cline{2-4}
                                                                & Unified        & \textbf{61.58} & 57.49(D)\\\cline{2-4} 
            \multicolumn{1}{|l|}{}                              & Divided         & 54.14 & 56.25(D)\\ \hline
        \end{tabular}
        \caption{Predictions of party affiliation from question utterances across hearing types, comparing unified and divided Congress settings. The majority class is our baseline. An annotation of (R) or (D) indicates that Republican or Democrat speakers have more utterances.}
        \label{tab:UniDivGovPartyResultsHT_BERT}
    \end{table}

    \paragraph{Differences across Hearing Types.} 
    Table \ref{tab:UniDivGovPartyResultsHT_BERT} summarizes how predictive performance varies by hearing type and political context when we use question utterances to infer the speaker's party affiliation. A central takeaway is that party affiliation is generally more linguistically legible than party standing across hearing types, suggesting that party-linked differences in language are more stable than standing-linked differences. Across hearing types, affiliation prediction shows the strongest improvement over baseline in general and oversight hearings under most conditions. A notable exception is oversight hearings under divided government, where the majority-class baseline exceeds classifier performance. Overall, these results are consistent with the idea that hearings with high visibility or explicit accountability functions create incentives for partisan messaging and framing that differ by party \citep{lee2009}, and with accounts emphasizing strategic use of hearings for oversight and political positioning \citep{mccubbins1984}. By contrast, authorization hearings exhibit weaker affiliation signal, suggesting that when hearings focus on technical or distributive policy work, partisan language may be less distinctive \citep{krehbiel1991,gilligan1990}. Results on predicting party standing across hearing types are provided in Appendix~\ref{app:SupResults}.

\section{Discussion and Summary} \label{sec:discussion}
Our results establish that questions in congressional hearings are not neutral vehicles for oversight but are infused with partisan and institutional markers. Partisan signals vary substantially across committees, hearing types, and government configurations, suggesting that language-based measures can detect where partisan conflict leaves strong imprints in questioning style and where it does not. The asymmetry between question-based and answer-based prediction underscores the distinct rhetorical roles of questioners and witnesses in congressional hearings. The failure of zero-shot LLM prompting to exceed baseline further suggests that partisan linguistic patterns in institutional dialogue are qualitatively different from those in the political texts on which LLMs are typically evaluated.

Our dataset and pipeline generalize beyond congressional hearings to structured interview-like settings including media interviews, judicial proceedings, and expert testimony, providing a framework for comparative study of questioning practices across institutions. Future work will extend the dataset to other democratic institutions, link linguistic patterns to substantive outcomes such as policy changes and media coverage, and more systematically analyze how unified vs. divided government shapes rhetorical strategies.

\section*{Limitations}
\paragraph{Data Coverage and Lag.} Official transcripts published by the Government Publishing Office (GPO) are typically available only 2 months to 2 years after a hearing is held, meaning the dataset will always lag behind current congressional activity; expansion to more recent sessions is planned as transcripts become available. Many earlier congressional hearings are available only as scanned documents requiring OCR, introducing potential transcription errors and limiting extension to archival sessions; addressing this is a direction for future work.

\paragraph{Metadata Quality.} Hearing-type metadata contains known classification errors, such as treaties occasionally misclassified as general hearings, which may affect analyses that condition on hearing type.

\paragraph{Interpretability.} While our BERT-based classifier achieves strong predictive performance, interpreting what linguistic features drive predictions remains challenging; our NELA-based analysis provides a partial remedy by identifying interpretable linguistic differences between parties. While we demonstrate that party affiliation is detectable from questioning style, our analysis does not fully characterize the specific strategies that distinguish partisan questioning; a deeper qualitative and quantitative analysis of how members from different parties structure questions, deploy rhetorical devices, and frame issues remains an important direction for future work.

\paragraph{LLM Evaluation.} Our evaluation of zero-shot LLM prompting reflects the models available at the time of the study; given the rapid pace of development in large language models, an extensive evaluation of LLMs at detecting party affiliation, a systematic comparison against lay and expert human annotators, and the development of a benchmark for this task, deserve a dedicated future study.

\paragraph{Scope and Generalizability.} Our substantive findings are specific to the U.S. congressional context and may not generalize directly to legislative questioning in other institutional settings, though our pipeline is designed to be extensible to such settings. The dataset covers only public hearings and does not include congressional meetings, where members deliberate without witnesses; question-answer dynamics in such settings may differ substantially.

\paragraph{Pipeline Robustness.} Our hybrid NER pipeline relies on the formatting conventions of GPO hearing transcripts and may require adaptation if transcript structure changes, such as those introduced by remote hearings or evolving publication standards; however, the framework is flexible enough to accommodate domain-specific NER implementations for other settings.

\paragraph{Party Representation.} The affiliation task treats Congress as effectively two-party, excluding Independent members; this simplification may become less accurate if third-party representation grows.

\paragraph{Open Questions.} Our findings raise broader questions including whether partisan linguistic patterns in hearings have intensified over time in tandem with polarization, whether questioning strategies vary by member seniority or electoral vulnerability, and whether the linguistic features identified here predict downstream outcomes such as media coverage or policy change. Connecting committee hearing proceedings to actual legislative outcomes and lawmaking remains a broader challenge, as the causal pathways between hearing language and policy are complex and difficult to trace. Although we do not provide methods or results to address these questions in this work, we expect that our dataset will enable such future investigations.

\section*{Ethical Considerations}
\paragraph{Data and Privacy.} The C-QUERI dataset is constructed exclusively from public congressional hearing transcripts published by the Government Publishing Office. All statements made in congressional hearings are part of the public record, and members of Congress and witnesses appear in their official capacities. No personally identifiable information beyond what is already in the public record is collected, stored, or released as part of this dataset. We do not include any data about private individuals.

\paragraph{Potential for Misuse.} Our dataset and methods are designed to advance scholarly understanding of legislative behavior and democratic accountability. However, we acknowledge that tools capable of detecting partisan signals in language could potentially be misused, for example to profile individuals based on their communication style or to generate targeted political messaging. We encourage users of the dataset and methods to consider the ethical implications of downstream applications and to adhere to responsible use guidelines. We also note a dual-use tension inherent in our work: while the ability to identify effective questioning strategies may benefit legislators and researchers seeking to improve the quality of oversight, the same methods could be used defensively by witnesses or their advisors to anticipate and prepare for partisan lines of questioning. We believe the transparency afforded by open publication of our methods and data is preferable to leaving such analysis to well-resourced private actors alone.

\paragraph{Bias and Representation.} Our analysis focuses on the two major U.S. political parties, which reflects the structure of Congress but excludes Independent members and third-party voices. Users of the dataset should be aware that models trained on this data may reflect and reinforce existing partisan categories. Additionally, our dataset spans a period of increasing political polarization, which may mean that models trained on it encode the specific rhetorical patterns of this era rather than more general properties of political discourse.

\paragraph{Broader Impact.} By making the dataset and pipeline publicly available, we aim to lower the barrier to entry for interdisciplinary research at the intersection of NLP and political science. We hope that this resource will support research on democratic accountability, legislative representation, and the health of democratic institutions, and that the framework we provide will be extended to study legislative questioning in other democratic contexts worldwide.

\bibliography{references}

\newpage
\appendix

\section{Supplementary Statistics for the Dataset} \label{app:Supp_Dataset_Statistics}
This section contains the supplementary statistics and related discussions.

\begin{figure}[h]
    \centering
    \begin{tabular}{c}
        \includegraphics[width=1\linewidth]{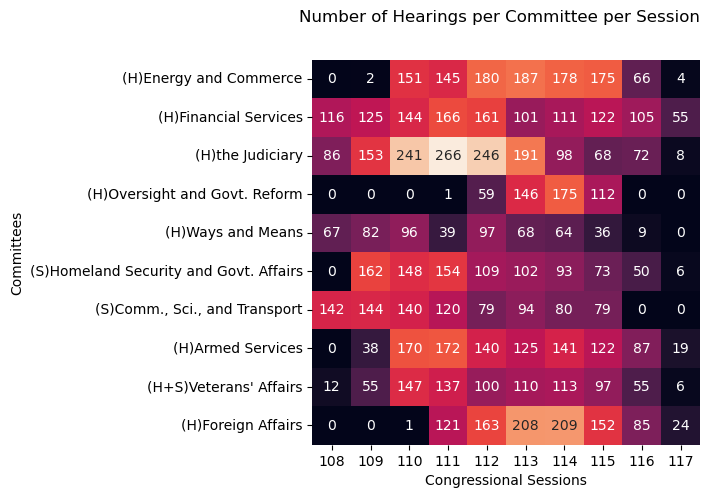} \\
        (a) \\ 
        \includegraphics[width=1\linewidth]{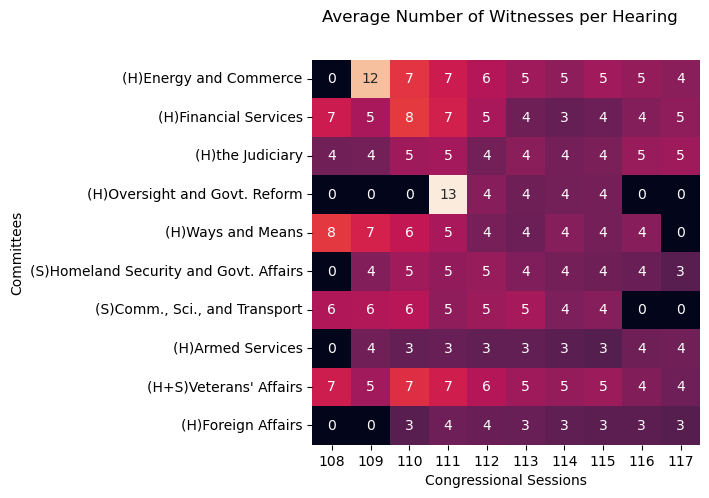} \\
        (b)
    \end{tabular}    
    \caption{The (a) number of hearings and (b) average number of witnesses by committee and session}
    \label{fig:hearingshm}
    \end{figure}

    Fig.\ref{fig:hearingshm} (a) illustrates the distribution of hearings across congressional sessions and the top ten most active committees. We notice some clear trends of congressional committees responding to domestic and global events. The House Committee on Financial Services which oversees the financial services industry in the United States shows a clear increase in the number of hearings in response to the 2007--2008 global financial crisis between the 110th and 112th congressional sessions which span the years 2007 to 2013. We also see a clear increase in the average number of witnesses called to hearing within the same period in Fig.\ref{fig:hearingshm} (b). A spike in the number of hearings by the Judicial committee in the 110th session coincides with growing concerns over domestic warrantless surveillance and wiretapping by law enforcement and national security agencies.

    \begin{figure}[h]
        \centering
        \begin{tabular}{c}
            \includegraphics[width=1\linewidth]{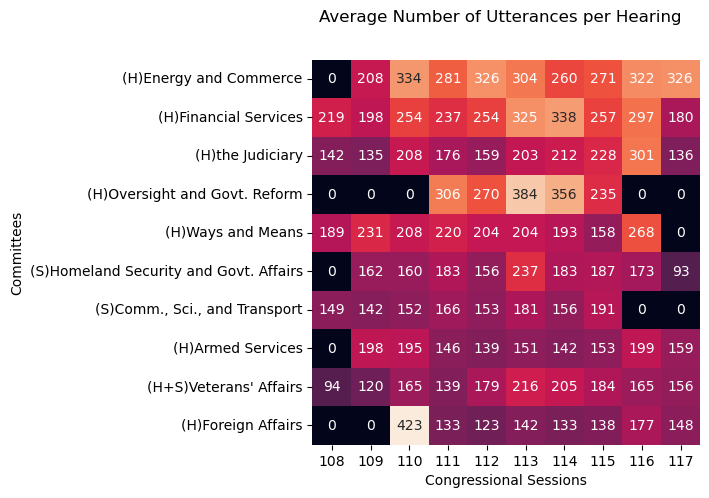} \\
            (a) \\
            \includegraphics[width=1\linewidth]{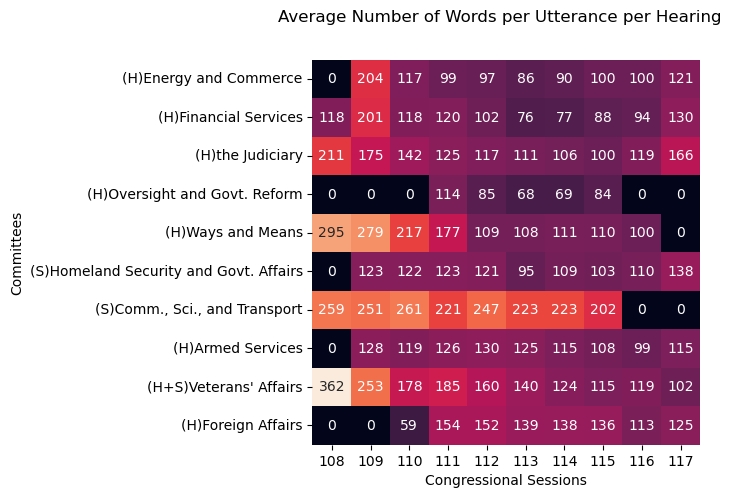} \\
            (b)
        \end{tabular}
        \caption{Heatmaps of (a) the number of utterances per hearing and (b) the average number of words per utterance, across committees and congressional sessions}
        \label{fig:utteranceshm}
    \end{figure}
        
\section{Identifying Questions and Answers}
\label{app:QnA}
This section discusses, in further detail, the datasets used for training the BERT classifier for the Question and Answer identification task discussed in Section~\ref{subsec:qna}.
\subsection{Datasets used for transfer learning}
\begin{figure}[H]
            \centering
            \includegraphics[width=1\linewidth]{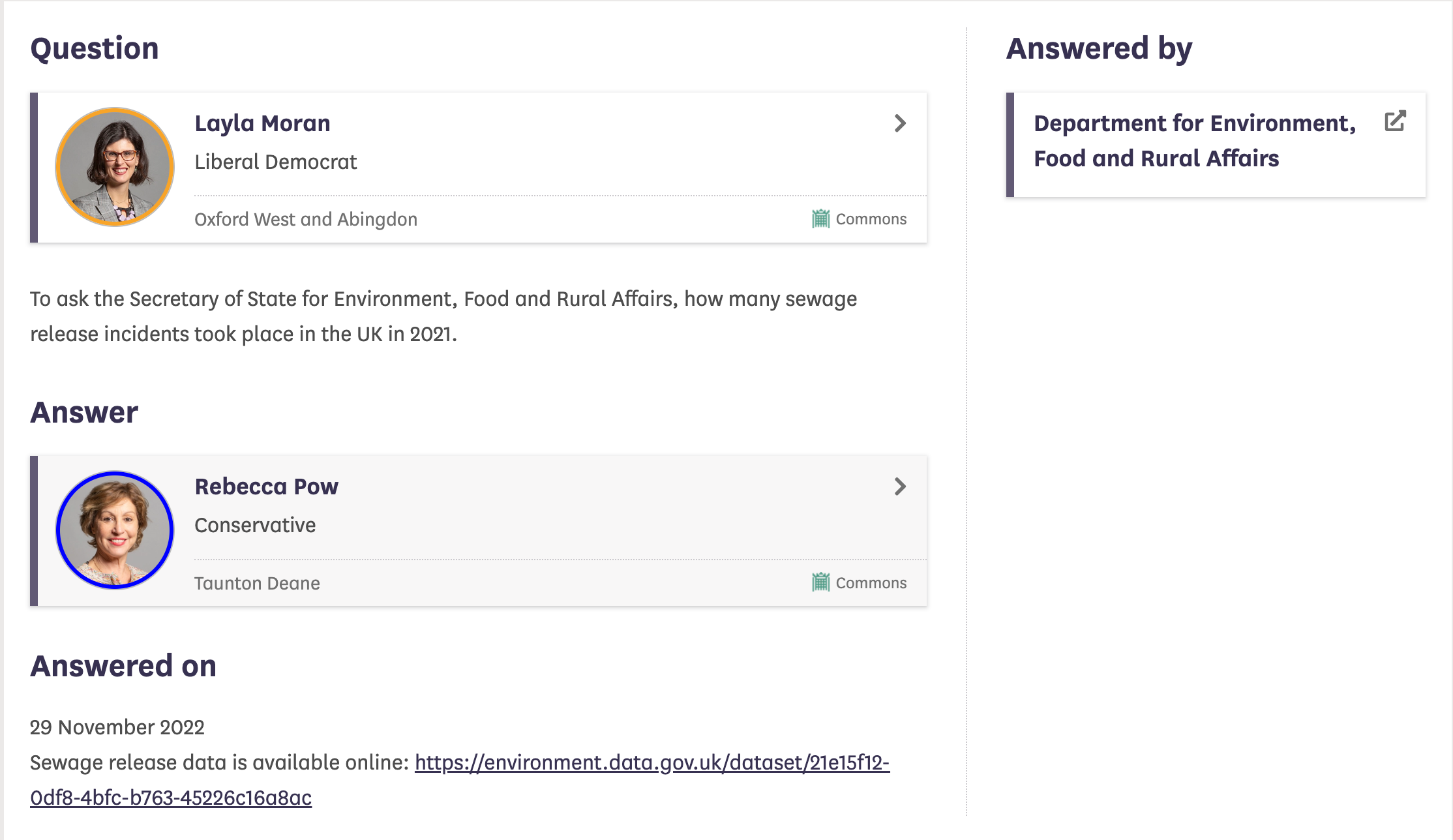}
            \caption{An example illustrating a question answered by a member of the U.K. Parliament in writing and the response from the concerned minister. Questions and answers are clearly annotated.}
            \label{fig:UKP}
        \end{figure}

    \paragraph{U.K. Parliamentary Question Hour}
        The U.K. Parliamentary written answered questions are an important aspect of the democratic process in the United Kingdom. Members of Parliament (M.P.s) can submit written questions to government ministers on any issue within their portfolio. These questions are typically answered in writing within a set timeframe and are publicly available online on the U.K. Parliament's official website. This provides an opportunity for M.P.s to hold ministers accountable for their actions, even when Parliament is not in session. This dataset also provides an excellent source for labeled examples of questions and answers. We have used 4688 such examples (2344 questions and 2344 answers) to test our question-answer prediction model. Figure \ref{fig:UKP} is an example of an answered question from the House of Commons.
        
    \paragraph{r/AskMeAnything}
        \begin{figure}[H]
            \centering
            \begin{tabular}{c}
            \includegraphics[width=1\linewidth]{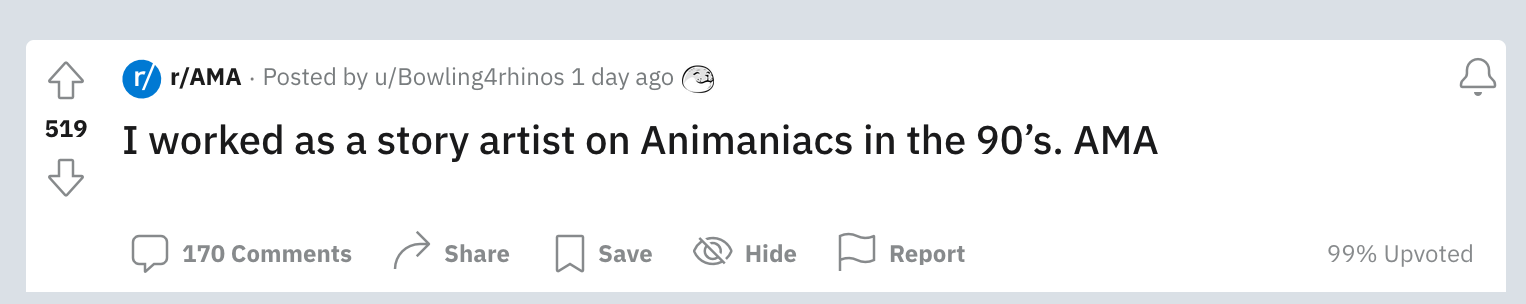} \\
            (a)\\
            \includegraphics[width=1\linewidth]{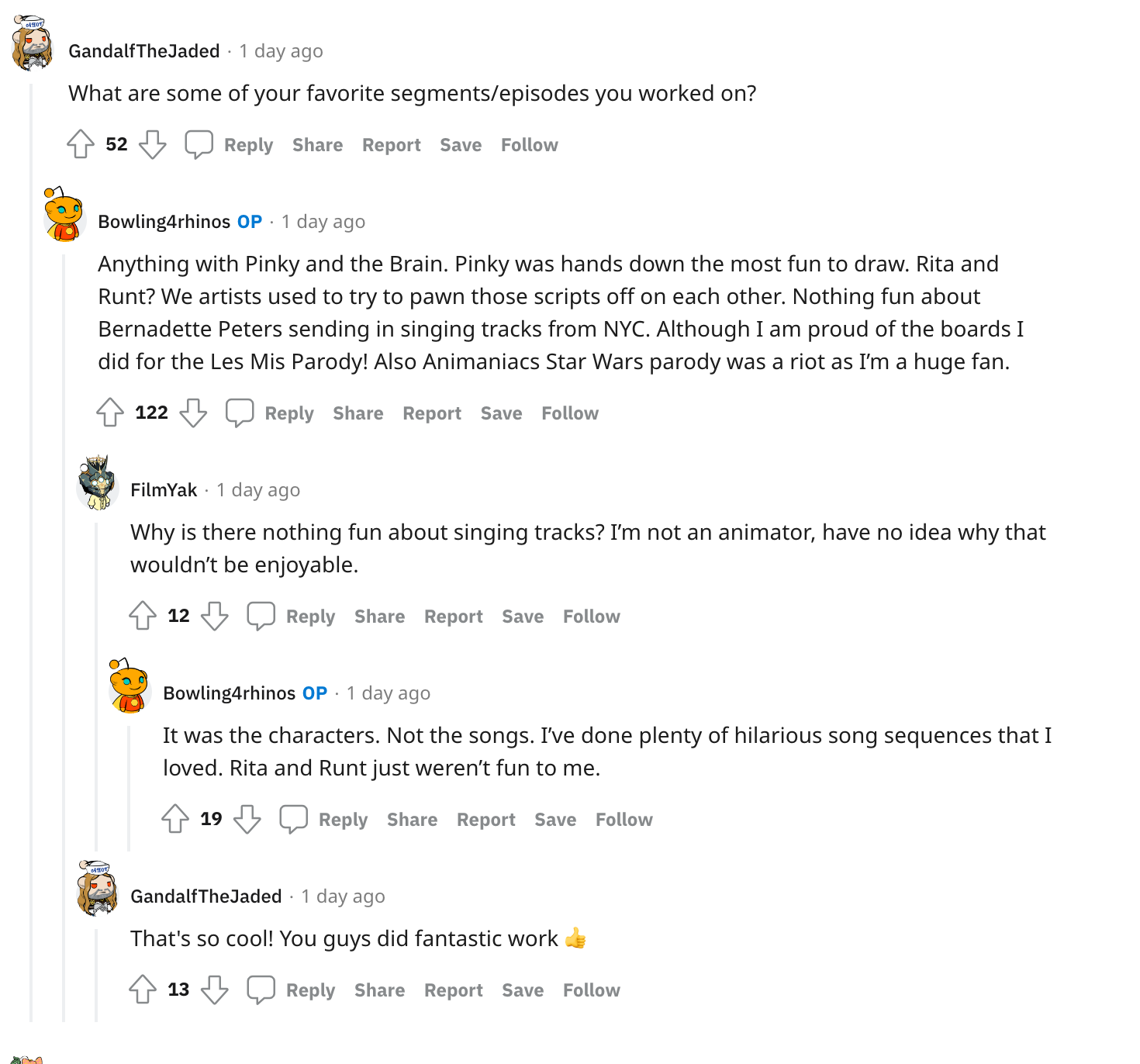} \\
            (b) \\
            \end{tabular}
            \caption{An Ask Me Anything (AMA) post in the subreddit r/AMA (a), and the subsequent comments on the post (b).}
            \label{fig:AMA}
        \end{figure}
        The subreddit AskMeAnything, also known as AMA, is a popular online platform on Reddit that allows users to host a question-answering session that allows other users to ask the host any question. Hosts may include celebrities, politicians, scientists, athletes, and everyday individuals with unique experiences or perspectives to share. AMA sessions are typically hosted in a question-answer format, where the person answering questions, also known as the ``host,'' responds to questions posted by users in real time. This makes the posts a very rich source of labeled question-answer examples. We construct a dataset with 121512 examples, 60756 questions, and 60756 answers as follows to train our questions-answer prediction model. Each post is an introduction where the expert introduces themselves; Figure \ref{fig:AMA} (a) is an example of such a post and Figure \ref{fig:AMA} (b) is an example of comments on the same post. The first level of comments on these posts are typically questions, and comments by the post creator in the second level of comments answer the questions. 

\section{Feature Descriptions from the NELA Toolkit}
This section contains details about the features from the NELA toolkit~\cite{horne_2020_NELA} refered to in Section~\ref{sec:PartyPrediction}. Table~\ref{table:Complexity_Description} contains the computations of the Complexity features, Table~\ref{table:Affect_Description} and Table~\ref{table:Bias_Description} contain the descriptions of the Affect and Bias features respectively.

\begin{table}[t]
    \centering
    \footnotesize
    \begin{tabular}{|l|p{7cm}|}
        \hline
        Feature & Description \\\hline
         ttr & Lexical Diversity also known as Type-Token Ratio\\\hline
         avgWlen & Average number of characters in a word\\\hline
         wCount & Number of words\\\hline
         FKGLvl & \textbf{Flesch–Kincaid grade level}: Standard readability measure computed by \[0.39*\dfrac{total words}{total sentences} + 11.8*\dfrac{total syllables}{total words} - 15.59\]\\\hline
         SmgIn & \textbf{Smog Index}: Standard readability measure computed by \[1.0430*\sqrt{\#polysyllables * \dfrac{30}{\#sentences}}+ 3.1291\]\\\hline
         CLIn & \textbf{Coleman–Liau index}: Standard readability measure computed by \[0.0588*L - 0.296*S -15.8\] where L = avg \# letters per 100 words and S = avg \# sentences per 100 words \\\hline
         lix & \textbf{LIX}: Standard readability measure computed by \[S+L\] where S = average sentence length and L = percentage of words with more than 6 letters. The scores usually range from 20 to 60. \\
         \hline
    \end{tabular}
    \caption{Description of the \bf{Complexity} category of NELA Features}
    \label{table:Complexity_Description}
\end{table}

\begin{table}[H]
    \centering
    \footnotesize
    \begin{tabular}{|lp{5cm}|}
        \hline
        Feature & Description \\
        \hline
         vneg & Negative sentiment score using Vadar Sentimet\\
         vneu & Neutral sentiment score using Vadar Sentiment \\
         vpos & Positive sentiment score using Vadar Sentiment \\
         wneg & Number of weak negative words\\
         wpos & Number of weak positive words\\
         wneu & Number of weak neutral words\\
         sneg & Number of strong negative words\\
         spos & Number of strong positive words\\
         sneu & Number of strong neutral words\\
         \hline
    \end{tabular}
    \caption{Description of the \bf{Affect} category of NELA Features}
    \label{table:Affect_Description}
\end{table}

\begin{table}[H]
    \centering
    \footnotesize
    \begin{tabular}{|lp{5cm}|}
        \hline
        Feature & Description \\
        \hline
         bias & The number of bias words\\
         assert & the number of assertive verbs\\
         facts & The number of factive verbs\\
         hedges & The number of hedge words\\
         implctv & The number of implicatives\\
         repVerb & Count of report verbs\\
         poWords & number of positive pinion words\\
         noWords & number of negative opinion words\\
        \hline
    \end{tabular}
    \caption{Description of the \bf{Bias} category of NELA Features}
    \label{table:Bias_Description}
\end{table}

\section{Statistical Testing with NELA Features}\label{app:c}

This appendix reports a detailed linguistic analysis based on two-sample Kolmogorov–Smirnov (K–S) tests of NELA features as discussed in Section~\ref{sec:PartyPrediction_result}. For each comparison, we sampled question utterances from members of different parties, with the null hypothesis that both samples were drawn from the same distribution.

\begin{figure*}[p]
        \centering
        \begin{subfigure}{.4\textwidth}
          \centering
          \includegraphics[width=1\linewidth]{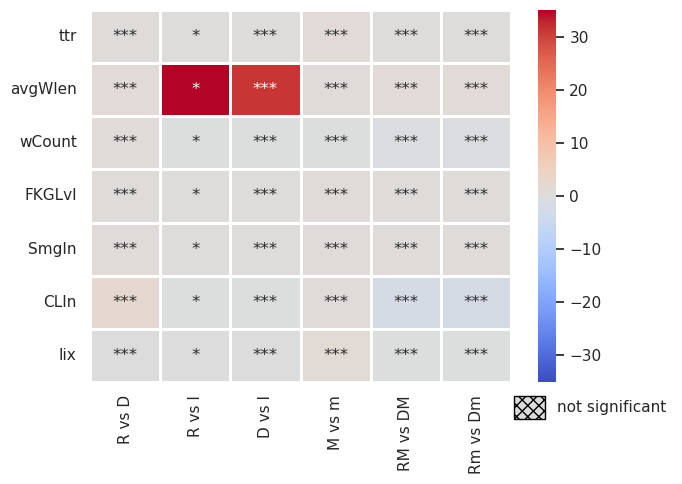}
          \caption{Statistical Testing of the Answer Utterances}
          \label{fig:A_Complexity}
        \end{subfigure}%
        \begin{subfigure}{.4\textwidth}
          \centering
          \includegraphics[width=1\linewidth]{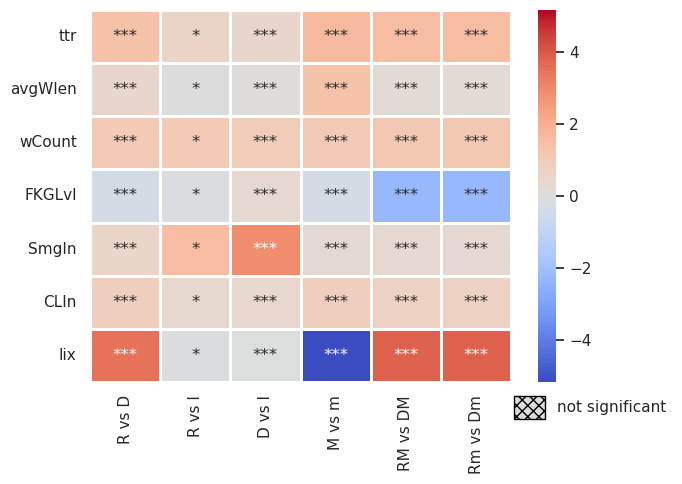}
          \caption{Statistical Testing of the Question Utterances}
          \label{fig:Q_Complexity}
        \end{subfigure}
        \caption{Statistical Testing of the Complexity NELA Features using two-sample Kolmogorov-Smirnov test. The hue of each cell signifies the ratio of the means of the two samples being studied; red  hue signifies that the left sample is larger where as the blue signifies that the right side sample has a larger mean. The number of asterisks `*' indicates the value of confidence for that statistical test. `***'$\implies p \in [0, 0.001)$, `**'$\implies p\in[0.001, 0.01)$, `*' $\implies p\in [0.01, 0.05).$ Differences that are not statistically significant are indicated with a cross-hatched pattern.}
        \label{fig:Complexity}
    \end{figure*}

\begin{figure*}[p]
        \centering
        \begin{subfigure}{.4\textwidth}
          \centering
          \includegraphics[width=1\linewidth]{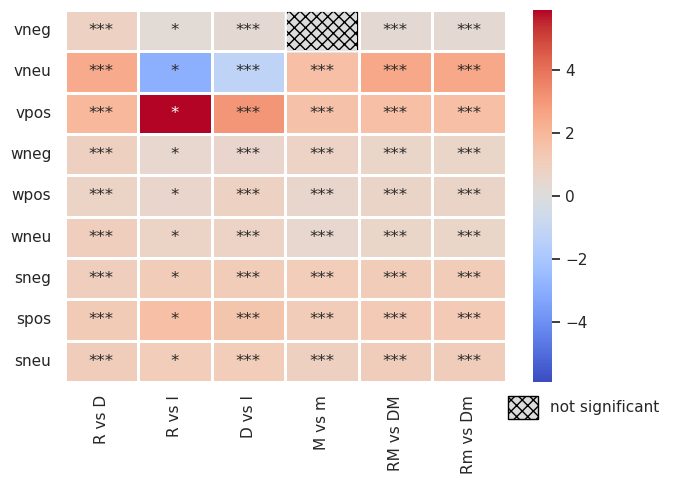}
          \caption{Statistical Testing of the Answer Utterances}
          \label{fig:A_Affect}
        \end{subfigure}%
        \begin{subfigure}{.4\textwidth}
          \centering
          \includegraphics[width=1\linewidth]{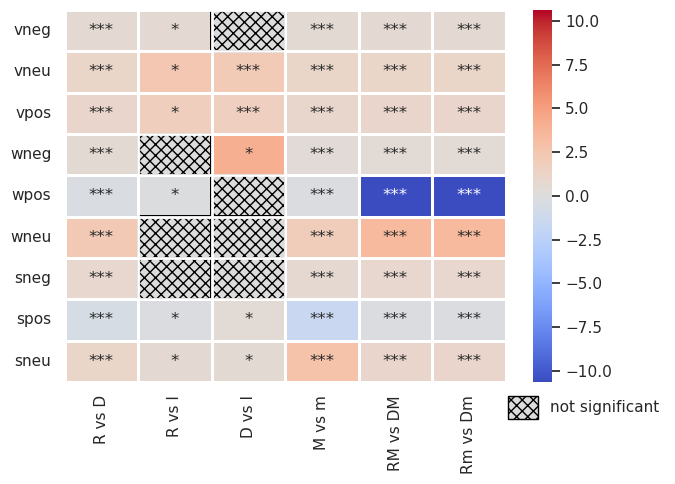}
          \caption{Statistical Testing of the Question Utterances}
          \label{fig:Q_Affect}
        \end{subfigure}
        \caption{Two-sample Kolmogorov-Smirnov test on the Affect NELA Features using. The hue of each cell signifies the ratio of the means of the two samples being studied; red  hue signifies that the left sample is larger where as the blue signifies that the right side sample has a larger mean. The number of asterisks `*' indicates the value of confidence for that statistical test. `***'$\implies p \in [0, 0.001)$, `**'$\implies p\in[0.001, 0.01)$, `*' $\implies p\in [0.01, 0.05)$. Differences that are not statistically significant are indicated with a cross-hatched pattern.}
        \label{fig:Affect}
    \end{figure*}

\begin{figure*}[p]
        \centering
        \begin{subfigure}{.4\textwidth}
          \centering
          \includegraphics[width=1\linewidth]{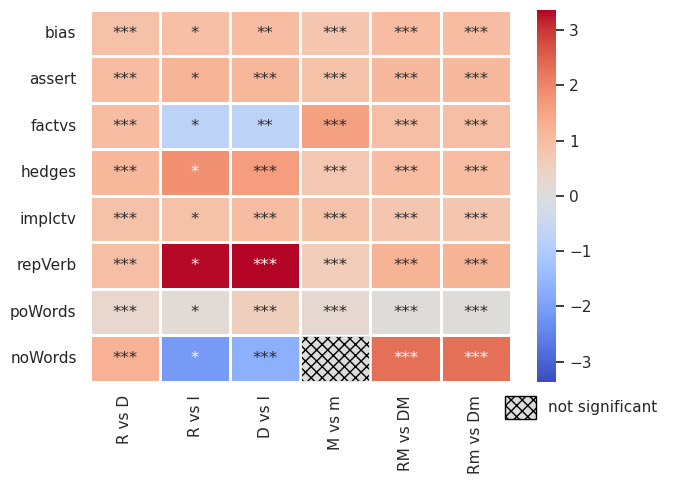}
          \caption{Statistical Testing of the Answer Utterances}
          \label{fig:A_Bias}
        \end{subfigure}%
        \begin{subfigure}{.4\textwidth}
          \centering
          \includegraphics[width=1\linewidth]{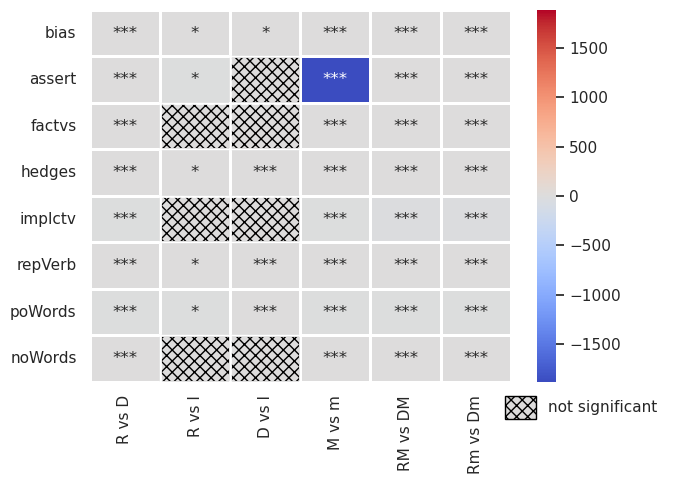}
          \caption{Statistical Testing of the Question Utterances}
          \label{fig:Q_Bias}
        \end{subfigure}
        \caption{Two-sample Kolmogorov-Smirnov test on the Bias NELA Features using. The hue of each cell signifies the ratio of the means of the two samples being studied; red  hue signifies that the left sample is larger where as the blue signifies that the right side sample has a larger mean. The number of asterisks `*' indicates the value of confidence for that statistical test. `***'$\implies p \in [0, 0.001)$, `**'$\implies p\in[0.001, 0.01)$, `*' $\implies p\in [0.01, 0.05).$ Differences that are not statistically significant are indicated with a cross-hatched pattern.}
        \label{fig:Bias}
    \end{figure*}

Our analysis reveals several patterns; on average, both Democrats and Republicans use more complex language than Independents (\Cref{fig:Complexity}). Democrats also employ more positive sentiment words, whereas Republicans rely more heavily on neutral sentiment (\Cref{fig:Bias}). Finally, Republicans make greater use of assertive and hedge words, while Democrats more often use implicative and positive opinion words (\Cref{fig:Affect}).

The hues of each cell represent the difference between the means of the two distributions under study. If the difference between the two distributions is not statistically significant, the cell has been hatched with the crossed pattern. Here, the R stands for Republican, D for Democrat, I for Independent, M for Majority party, m for Minority party, R.M. for Republican in Majority, and D.M. for Democrat in Majority.

    
\section{Hyperparameter Search} \label{app:Gridsearch}
    We discuss the hyperparameters that we have used for our gridsearches for the different models in Tables \ref{tab:NELA_hyperparameters}, \ref{tab:SVC_hyperparameters}, \ref{tab:LogisticRegression_hyperparameters}, and \ref{tab:BERT_hyperparameters}.

    \begin{table}[H]
        \footnotesize
        \centering
        \begin{tabular}{|l|p{4cm}|}
            \hline
             \textbf{Hyperparameter} & \textbf{Search Grid}  \\\hline
              n\_estimators & {10, 20, ..., 100} (step of 10)\\\hline
              criterion & {gini, entropy} \\\hline
              max\_depth & {50, 100, ..., 500} (step of 50)\\\hline
              min\_samples\_split & {0.0001, 0.005, 0.001, 0.05, 0.01, 0.5, 0.2, 0.1 }\\\hline
        \end{tabular}
        \caption{Random Forest Hyperparameters}
        \label{tab:NELA_hyperparameters}
    \end{table}
    
    \begin{table}[H]
        \footnotesize
        \centering
        \begin{tabular}{|l|p{4cm}|} \hline
        \textbf{Hyperparameter} & \textbf{Search Grid}  \\\hline
            C & {0.001, 0.01, 0.1, 1, 10, 100}  \\\hline
            penalty & {l1, l2} \\\hline
            solver & {liblinear, saga} \\\hline
        \end{tabular}
        \caption{Hyperparameters for Logistic Regression}
        \label{tab:LogisticRegression_hyperparameters}
    \end{table}

    \begin{table}[H]
        \footnotesize
        \centering
        \begin{tabular}{|l|p{4cm}|}
            \hline
            \textbf{Hyperparameter} & \textbf{Search Grid}  \\\hline
             C & {0.1, 1, 10, 100} \\\hline
             gamma & {0.001, 0.01, 0.1, 1} \\\hline
             kernel & {rbf,sigmoid,linear}  \\\hline
        \end{tabular}
        \caption{Hyperparameters for Support Vector Classifier}
        \label{tab:SVC_hyperparameters}
    \end{table}
    
    \begin{table}[H]
        \footnotesize
        \centering
        \begin{tabular}{|l|p{4cm}|}
            \hline
             \textbf{Hyperparameter} & \textbf{Search Grid}  \\\hline
              Initial Learning Rate & {1e-5,1e-7} \\\hline
              \# Epochs & {5, 10, 15, 20} \\\hline
              Batch Sizes & {2, 32, 64, 128} \\\hline
        \end{tabular}
        \caption{BERT Hyperparameters}
        \label{tab:BERT_hyperparameters}
    \end{table}

\section{LLM Model Versions and Configurations} \label{app:LLM_Config}
\begin{figure*}[htp]
        \centering
        \includegraphics[width=.9\linewidth]{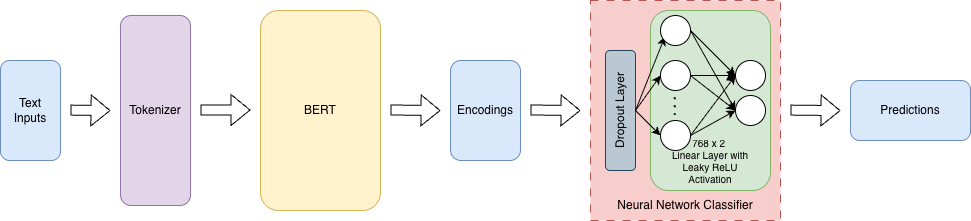}
        \caption{Architecture of the BERT classifier model}
        \label{fig:BERT}
    \end{figure*}

    In this section we discuss the large language models (LLMs) that we have used in our experiments. The identifiers in parentheses reflect the version strings:

    \begin{itemize}[noitemsep]
        \item BERT Classifier (see Figure~\ref{fig:BERT}) (bert-base-cased, fine-tuned on the different tasks)
        \item LLaMA-3-70B (llama3:70b)
        \item DeepSeek-R1-70B (deepseek-r1:70b)
        \item Qwen-3-32B (qwen3:32b)
        \item Gemma-3-27B (gemma3:27b)
        \item Mistral (mistral:latest)
    \end{itemize}
    For each prediction task, we queried the large language models using zero-shot prompts designed to elicit the target label (party affiliation or party standing) directly from the text of the question utterance. The speaker’s name, typically present at the start of each utterance, was removed to avoid trivial cues. Prompts were phrased in a standardized format across all models to ensure consistency, with variations only in the input text corresponding to each utterance. The exact wording of the prompts is provided in Table~\ref{tab:LLM_prompts}, for reproducibility.

\begin{table}[h]
    \footnotesize
    \centering
    \begin{tabular}{|p{1.1cm}|p{1.1cm}|p{1.5cm}|p{1.5cm}|}
    \hline
    \multicolumn{4}{|p{.95\linewidth}|}{\textbf{Prompt:}``What follows is a \{type\_text\} in a congressional hearing: \{utterance\_text\} The question was asked by a person who is a member of a congressional committee, and whose party affiliation is either Democrat, Independent, or Republican. Based on the \{type\_text\_2\} above, what is the party affiliation of the person who asked the question? Answer with either D for Democrat, I for Independent, or R for Republican. Do not explain.''}\\\hline
    \multicolumn{1}{|c|}{\multirow{2}{*}{\bf Variable name}} & \multicolumn{3}{c|}{\bf Text according to utterance type} \\ \cline{2-4}
    & \multicolumn{1}{c|}{\textbf{Question}} & \multicolumn{1}{c|}{\textbf{Answer}} & \multicolumn{1}{c|}{\textbf{Both}} \\\hline
    ``type\_text''  & ``question that has been asked'' & ``response to a question asked'' & ``question and its answer'' \\ \hline
     ``utterance\_text'' & ``Question:'' followed by the text of the utterance &  `` Answer:'' followed by the text of the utterance & ``Question:'' followed by the text of the question and then ``Answer:'' followed by the text of the answer \\\hline
     ``type\_text\_2'' & ``question'' & ``answer'' & ``question and answer''\\\hline
    \end{tabular}
    \caption{Prompts used for zero-shot predictions of party affiliation for different utterance types, standardized across all LLMs.}
    \label{tab:LLM_prompts}
\end{table}

\section{GPU and Environment Configurations} \label{app:Env}
All LLM experiments were run on an NVIDIA RTX 6000 Ada Generation GPU with 49,140 MiB of memory. Inference was performed in a zero-shot setting without fine-tuning, using various batch-sizes across the models. Experiments were conducted in a Python 3.10 environment with PyTorch 2.2.1 and Hugging Face Transformers 4.41.2, running on CUDA 12.1 with cuDNN 8.9.

\section{Full Results (Supplementary)} \label{app:SupResults}
    This appendix provides supplementary tables containing additional results and related discussion that did not fit in Section~\ref{sec:PartyPrediction} and Section~\ref{sec:PartyPrediction_result}. These include extended breakdowns by committee, hearing type, and linguistic feature, complementing the analyses presented in the paper.

    \noindent{\bf Predicting standing.} The results in Table \ref{tab:UniDivGovMajorityResultsHT}, on predicting the party standing of the questioner, provide several insights for scholars interested in studying politics. First, we observe from the baselines that majority party members (M) often dominate question utterances, reflecting the procedural advantages of majority status. Committee chairs control agendas, allocate time, and structure witness panels \cite{coxmccubbins2005}. Second, BERT outperforms the baseline in nomination hearings under unified government and Republican presidencies, where the Democrat minority generates the most questions. This indicates that minority questioners adopt linguistically distinctive strategies when challenging nominees, consistent with accounts of confirmation proceedings serving as an arena for opposition messaging and executive scrutiny \citep{binder2009}. We note that the high overall accuracy in oversight and field hearings largely reflects class imbalance, suggesting that the content of questions may be less distinguishable despite different participation rates in monitoring contexts, perhaps because both parties engage in credit-claiming or position-taking with similar rhetorical templates \citep{mayhew1974congress}.
    
    \begin{figure}[H]
        \centering
        \includegraphics[width=0.9\linewidth]{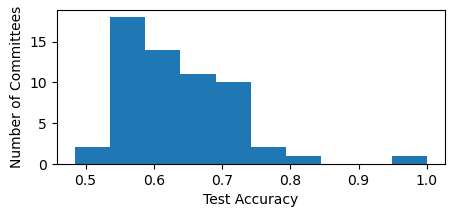}
        \caption{Histogram of test accuracies of the Random Forest Classifier on the NELA features across all committees for party affiliation.}
        \label{fig:Committee_Accuracy}
    \end{figure}
    
    \begin{table}[htp]
        \footnotesize
        \centering
        \begin{tabular}{|l|r|r|r|}
            \hline
            \textbf{Hearing Type} & \textbf{\#Hearings} & \textbf{\#Utterances} & \textbf{\#Questions}\\\hline
            General & 16287 & 2234567 & 579282\\\hline
            Feild & 467 & 46933 & 11962\\\hline
            Oversight & 330 & 51581 & 14075\\\hline
            Authorization & 479 & 58738 & 15381\\\hline
            Nomination & 357 & 34360 & 8179\\\hline
            Treaty 		& 1 	& 88 	& 22\\\hline
            Markup 		& 8 	& 2001 	& 16\\\hline
        \end{tabular}
        \caption{The number of hearings, utterances and questions per hearing type.}
        \label{tab:HearingTypes_Number}
    \end{table}
    
    \begin{table}[htp]
        \centering
        \footnotesize
        \begin{tabular}{|c|c|c|p{0.4\linewidth}|}
            \hline
            \textbf{NELA} & \textbf{Baseline} & \textbf{Session} & \textbf{Committee} \\
            \hline
             100.00 & 51.67(R) & 109 & (H)Committee on Financial Services and Committee on Resources \\ \hline
             73.78 & 61.39(D) & 113 & (S)Commerce, Science, and Transportation \\ \hline 
             76.00 & 55.17(D) & 110 & (S)Education and Labor and Committee on Health, Education, Labor, and Pensions \\ \hline
             96.00 & 58.33(D) & 113 & (H)Committee on Armed Services Meeting Jointly with Subcommittee on Asia and the Pacific\\ \hline
             76.36 & 50.00 & 117 & (H)House Administration \\ \hline
        \end{tabular}
        \caption{The top 5 combinations of Committees and Congressional session on which random forest classifiers using NELA features have the highest prediction accuracy in identifying party affiliations on question utterances. The Baseline column represents the majority-classifier.}
        \label{tab:ParQNELA}
    \end{table}

    \begin{table}[p]
        \footnotesize
        \centering
        \begin{tabular}{|l|c|c|c|}
            \hline
            \textbf{Model} & \textbf{Question} & \textbf{Answer} & \textbf{Both} \\ \hline \hline
            \multicolumn{1}{|p{2.5cm}|}{Llama}        & 55.91 & 54.67 & 58.37 \\ \hline
            \multicolumn{1}{|p{2.5cm}|}{Gemma}        & 55.63 & 52.81 & 58.50 \\ \hline
            \multicolumn{1}{|p{2.5cm}|}{Deepseek}     & 55.62 & 52.84 & 60.78 \\ \hline
            \multicolumn{1}{|p{2.5cm}|}{Qwen}         & 50.02 & 48.61 & 55.20 \\ \hline
            \multicolumn{1}{|p{2.5cm}|}{Mistral}      & 54.01 & 50.24 & 56.14\\ \hline
            \multicolumn{1}{|p{2.5cm}|}{LLM Baseline} & 52.12 & 52.12 & 52.12 \\ \hline
        \end{tabular}
        \caption{Results for identifying the questioner's party affiliation when the input is the question, the answer in response to a question or both question and answer text using zero-shot prompting.}
        \label{tab:Party_Results_LLM}
    \end{table}
    
    \begin{table*}[p]
        \centering
        \footnotesize
        \begin{tabular}{|c|c|c|c|c|c|l|}
            \hline
            \multicolumn{2}{|c|}{\textbf{All}} & \multicolumn{2}{c|}{\textbf{No Stop Words}} & \multicolumn{2}{c|}{\textbf{No Speaker}} & \multicolumn{1}{l|}{\multirow{2}{*}{\textbf{Committee on}}} \\\cline{1-6}
            \textbf{BERT} & \textbf{Base} & \textbf{BERT} & \textbf{Base} & \textbf{BERT} & \textbf{Base}  &\\\hline
             98.31 & 54.21 & 98.40 & 54.20 & 64.48 & 54.75 & (H)Energy and Commerce\\ \hline
             98.73 & 56.72 & 98.72 & 56.67 & 65.39 & 56.40 & (H)Financial Services\\ \hline 
             97.88 & 54.88 & 97.87 & 55.05 & 62.54 & 55.19 & (H+S)the Judiciary\\ \hline
             98.80 & 69.48 & 98.84 & 69.42 & 72.48 & 70.21 & (H)Oversight and Government Reform\\ \hline
             95.06 & 54.69 & 94.91 & 54.92 & 64.69 & 66.47 & (H)Ways and Means\\ \hline
             98.43 & 63.71 & 98.18 & 63.59 & 67.13 & 63.99 & (S)Homeland Security and Governmental Affairs\\ \hline
             98.94 & 61.34 & 99.07 & 61.43 & 60.95 & 60.54 & (S)Commerce, Science, and Transportation\\ \hline
             95.82 & 51.75 & 95.68 & 51.62 & 59.27 & 51.20 & (H)Armed Services \\ \hline
             98.82 & 50.45 & 99.03 & 50.10 & 61.04 & 50.32 & (H+S)Veterans' Affairs \\ \hline
             97.10 & 52.91 & 97.01 & 53.38 & 60.00 & 50.80 & (H)Foreign Affairs \\ \hline
        \end{tabular}
        \caption{Question utterance party affiliation prediction accuracy by session and committee obtained by training a BERT-based classifier}
        \label{tab:ParQBERT_Utterances}
    \end{table*}

    \begin{table*}[p]
        \centering
        \footnotesize
        \begin{tabular}{|c|c|c|c|c|c|l|}
            \hline
            \textbf{Llama}  & \textbf{Gemma} & \textbf{Deepseek} & \textbf{Qwen} & \textbf{Mistral} & \textbf{Baseline} & \textbf{Committee on} \\ \hline
            \textbf{56.38}  & \textbf{58.60} & \textbf{56.82}    & 50.51         & \textbf{56.04}   & 54.36(R) & (H)Energy and Commerce\\ \hline
            52.09           & \textbf{58.05} & \textbf{58.38}    & 51.66         & 55.76            & 57.31(R) & (H)Financial Services\\ \hline 
            \textbf{60.41}  & 52.45          & \textbf{56.45}    & 52.72         & 51.85            & 57.30(D) & (H+S)the Judiciary\\ \hline
            52.01           & 66.85          & 59.08             & 44.53         & 63.18            & 70.85(R) & (H)Oversight and Govt. Reform\\ \hline
            45.95           & 62.16          & 54.55             & 60.00         & \textbf{71.43}   & 65.00(R) & (H)Ways and Means\\ \hline
            54.53           & 44.12          & 47.90             & 43.23         & 36.89            & 60.47(D) & (S)Homeland Security and Govt. Affairs\\ \hline
            56.92           & 51.36          & 54.69             & 51.26         & 52.55            & 57.64(D) & (S)Comm, Sci, and Transportation\\ \hline
            \textbf{59.28}  & \textbf{51.89} & \textbf{56.79}    & 50.54         & \textbf{51.49}   & 50.77(R) & (H)Armed Services \\ \hline
            50.00           & 47.40          & \textbf{52.63}    & 45.97         & 49.18            & 51.18(D) & (H+S)Veterans' Affairs \\ \hline
            55.25  & \textbf{56.91} & 54.31    & 49.58           & 53.65   & 56.00(R) & Foreign Affairs \\ \hline
             0.86           &  0.63          &  5.21             &  3.24         &  2.72            & - & Not Answered by LLM\\ \hline
        \end{tabular}
        \caption{Zero-shot predictions of questioner party affiliation from question utterances (with the speaker's name removed), across the ten committees with the most utterances, using various LLMs. The Base column reports the majority-class baseline.}
        \label{tab:ParQLLM}
    \end{table*}

    \begin{table*}[p]
            \centering
            \footnotesize
            \setlength\tabcolsep{3pt}
            \begin{tabular}{|l|l|cccccc|}
                \hline
                \textbf{Committee on} & \textbf{Government}       & \textbf{Llama}  & \textbf{Gemma} & \textbf{Deepseek} & \textbf{Qwen} & \textbf{Mistral} & \textbf{Base} \\ 
                \hline
                \multicolumn{1}{|l|}{\multirow{2}{*}{(H) \textbf{Energy and Commerce}}}  
                                                        & Unified & \textbf{59.50}    & 49.38          & \textbf{55.23}  & 51.27         & 49.21            & 53.89(D)\\ 
                \cline{2-8}
                \multicolumn{1}{|l|}{}                  & Divided & 55.12             & \textbf{62.34} & 57.11 & 50.19         & \textbf{61.54}   & 57.71(R)\\
                \hline
                \multicolumn{1}{|l|}{\multirow{2}{*}{(H) \textbf{Financial Services}}}    
                    & Unified & 51.66             & 53.05          &  54.84         & 48.06         & 53.49            & 56.90(R)\\
                \cline{2-8}
                \multicolumn{1}{|l|}{}                  & Divided & 53.70             & \textbf{62.24} &  \textbf{62.08}  & 55.38         & 54.55            & 57.89(R)\\ 
                \hline
                \multicolumn{1}{|l|}{\multirow{2}{*}{(H+S) \textbf{the Judiciary}}}
                    & Unified & \textbf{57.73}    & 54.11          &  \textbf{56.58} & 52.82         & 53.82            & 56.51(D)\\
                \cline{2-8}
                \multicolumn{1}{|l|}{}                  & Divided & \textbf{62.34}    & 51.46          & 56.74  & 52.46         & \textbf{68.44}   & 58.03(D)\\
                \hline
                \multicolumn{1}{|l|}{\multirow{2}{*}{(H) \textbf{Oversight and Gov Reform}}}
                    & Unified & 47.52             & 63.37          &   57.61    & 48.48         & 70.41            & 71.03(R)\\
                \cline{2-8}
                \multicolumn{1}{|l|}{}                  & Divided & 53.02             & 67.63          & 61.34       & 43.62         & 33.33            & 70.82(R)\\
                \hline
                \multicolumn{1}{|l|}{\multirow{2}{*}{(S) \textbf{Homel Sec and Gov Affairs}}}   
                    & Unified & 52.29             & 44.70          &  43.28  & 40.67         & 35.41            & 59.66(D)\\
                \cline{2-8}
                \multicolumn{1}{|l|}{}                  & Divided & 56.42             & 43.63          &  48.36   & 45.38         & 16.16            & 61.13(D)\\
                \hline
                \multicolumn{1}{|l|}{\multirow{2}{*}{(S) \textbf{Comm, Sci, and Transports}}}  
                    & Unified & 55.25             & 46.11          &   56.47    & 46.89         & 49.15            & 56.92(D)\\
                \cline{2-8}
                \multicolumn{1}{|l|}{}                  & Divided & 60.53             & 54.61          & 51.35  & 52.67         & \textbf{100.00}  & 64.71(D)\\
                \hline
                \multicolumn{1}{|l|}{\multirow{2}{*}{(H) \textbf{Armed Services}}}   
                    & Unified & \textbf{64.33}    & 50.63          &   50.34    & 50.33         & 50.32            & 55.43(D)\\
                \cline{2-8}
                \multicolumn{1}{|l|}{}                  & Divided & \textbf{56.78}    & 52.52          &  \textbf{54.00}   & 50.64         & 40.00            & 53.94(R)\\
                \hline
                \multicolumn{1}{|l|}{\multirow{2}{*}{(H+S) \textbf{Veterans' Affairs}}}   
                    & Unified & 59.26             & 48.75          & 48.65   & 50.00         & 45.57            & 64.71(D)\\
                \cline{2-8}
                \multicolumn{1}{|l|}{}                  & Divided & 46.33             & 47.95          &  49.52     & 43.66         & 42.86            & 51.84(R)\\
                \hline
                \multicolumn{1}{|l|}{\multirow{2}{*}{(H) \textbf{Foreign Affairs}}}   
                    & Unified & \textbf{55.06}    & 46.07          &  \textbf{54.02}  & 51.69         & 48.86            & 53.00(D)\\
                \cline{2-8}
                \multicolumn{1}{|l|}{}                  & Divided & 55.31             & \textbf{60.44} &   53.38    & 48.86         & 50.00            & 59.00(R)\\
                \hline
            \end{tabular}
            \caption{Predictions of party affiliations (Republican, Democrat or Independent) from question utterances across hearing types, comparing unified and divided Congress settings. The Base column reports the majority-class baseline.(R) denotes that speakers from the Republican party have the majority number of utterances, (D) denotes Democratic speakers hold the majority class in the dataset.}
            \label{tab:UniDivGovPartyResultsC_LLM}
        \end{table*}

    \begin{table*}[p]
            \centering
            \footnotesize
            \setlength\tabcolsep{3pt}
            \begin{tabular}{ll|r|r|r|r|r|r|}
                \hline
                \multicolumn{2}{|c|}{\multirow{2}{*}{Hearing Type}} & \multicolumn{2}{c|}{\bf All} & \multicolumn{2}{c|}{\bf Democratic Presid} &\multicolumn{2}{c|}{\bf Republican Presidency}\\ 
                \cline{3-8}
                \multicolumn{2}{|l|}{} & Majority & Minority & Majority & Minority & Majority & Minority \\
                \hline
                \multicolumn{1}{|l|}{\multirow{2}{*}{\bf All}}  & Unified & 128189 & 85670 & 50065 & 28057 & 78124 &57613\\
                \cline{2-8}
                \multicolumn{1}{|l|}{}                              & Divided & 224285 & 136869 & 197205 & 109798 & 27080 &27071\\
                \hline
                \multicolumn{1}{|l|}{\multirow{2}{*}{\bf General}}  & Unified & 116560 & 78621 & 46011 & 26064 & 70549 &52557 \\
                \cline{2-8}
                \multicolumn{1}{|l|}{}                              & Divided & 210931 & 129141 & 185003 & 103790 & 25928 &25351\\
                \hline
                \multicolumn{1}{|l|}{\multirow{2}{*}{\bf Field}}    & Unified & 2582 & 1187 & 802 & 240 & 1780 &947\\ 
                \cline{2-8}
                \multicolumn{1}{|l|}{}                              & Divided & 4066 & 1433 & 3710 & 1059 & 356 &374\\
                \hline
                \multicolumn{1}{|l|}{\multirow{2}{*}{\bf Oversight}}& Unified & 4008 & 1533 & 1455 & 612 & 2553 &921\\ 
                \cline{2-8}
                \multicolumn{1}{|l|}{}                              & Divided & 2194 & 977 & 1729 & 806 & 465 &171\\ 
                \hline
                \multicolumn{1}{|l|}{\multirow{2}{*}{\bf Authorization}}& Unified & 2630 & 2053 & 1035 & 785 & 1595 &1268\\ 
                \cline{2-8}
                \multicolumn{1}{|l|}{}                              & Divided & 5827 & 3792 & 5496 & 3313 & 331 &479\\ 
                \hline
                \multicolumn{1}{|l|}{\multirow{2}{*}{\bf Nomination}}& Unified & 2409 & 2276 & 762 & 356 & 1647 &1920\\ 
                \cline{2-8}
                \multicolumn{1}{|l|}{}                              & Divided & 1252 & 1507 & 1252 & 811 & 0 &696\\ 
                \hline
            \end{tabular}
            \caption{Number of question utterances from members of the party holding the majority/minority in the respective chamber under a unified or divided government for all utterances, utterances with democratic presidency, and utterances with republican presidency.}
            \label{tab:HearingTypes_PartyStanding}
        \end{table*}

        \begin{table*}[p]
            \centering
            \footnotesize
            \setlength\tabcolsep{3pt}
            \begin{tabular}{ll|r|r|r|r|r|r|}
                \hline
                \multicolumn{2}{|c|}{\multirow{2}{*}{\bf Hearing Type}} & \multicolumn{2}{c|}{\bf All} & \multicolumn{2}{c|}{\bf Democratic Presidency} &\multicolumn{2}{c|}{\bf Republican Presidency}\\ 
                \cline{3-8}
                \multicolumn{2}{|l|}{} & Democrat & Republican & Democrat & Republican & Democrat & Republican \\
                \hline
                \multicolumn{1}{|l|}{\multirow{2}{*}{\bf All}}  & Unified & 107678 & 106181 & 50065 & 28057 & 57613 &78124\\
                \cline{2-8}
                \multicolumn{1}{|l|}{}                              & Divided & 173633 & 187521 & 141800 & 165203 & 31833 &22318\\
                \hline
                \multicolumn{1}{|l|}{\multirow{2}{*}{\bf General}}  & Unified & 98568 & 96613 & 46011 & 26064 & 52557 &70549\\
                \cline{2-8}
                \multicolumn{1}{|l|}{}                              & Divided & 162430 & 177642 & 132269 & 156524 & 30161 &21118\\
                \hline
                \multicolumn{1}{|l|}{\multirow{2}{*}{\bf Field}}    & Unified & 1749 & 2020 & 802 & 240 & 947 &1780\\
                \cline{2-8}
                \multicolumn{1}{|l|}{}                              & Divided & 2737 & 2762 & 2330 & 2439 & 407 &323\\
                \hline
                \multicolumn{1}{|l|}{\multirow{2}{*}{\bf Oversight}}& Unified & 2376 & 3165 & 1455 & 612 & 921 &2553\\
                \cline{2-8}
                \multicolumn{1}{|l|}{}                              & Divided & 2209 & 962 & 1729 & 806 & 480 &156\\
                \hline
                \multicolumn{1}{|l|}{\multirow{2}{*}{\bf Authorization}}& Unified & 2303 & 2380 & 1035 & 785 & 1268 &1595\\
                \cline{2-8}
                \multicolumn{1}{|l|}{}                              & Divided & 4684 & 4935 & 4219 & 4590 & 465 &345\\ 
                \hline
                \multicolumn{1}{|l|}{\multirow{2}{*}{\bf Nomination}}& Unified & 2682 & 2003 & 762 & 356 & 1920 &1647\\ 
                \cline{2-8}
                \multicolumn{1}{|l|}{}                              & Divided & 1548 & 1211 & 1228 & 835 & 320 &376\\
                \hline
            \end{tabular}
            \caption{Number of question utterances from members of the democratic or republican parties under a unified or divided government for all utterances, utterances with democratic presidency, and utterances with republican presidency.}
            \label{tab:HearingTypes_PartyAffiliation_questions}
        \end{table*}

    \begin{table*}[h]
            \centering
            \footnotesize
            \setlength\tabcolsep{3pt}
            \begin{tabular}{|l|c|cc|cc|cc|}
                \hline
                \multirow{2}{2cm}{\bf Hearing Type} & \multirow{2}{2cm}{\bf Government} & \multicolumn{2}{c|}{{\bf All}} & \multicolumn{2}{c|}{{\bf Democrat President}} & \multicolumn{2}{c|}{\bf Republican President} \\ \cline{3-8}
                & & {\bf BERT} & {\bf Base} & {\bf BERT} &{\bf Base} & {\bf BERT} & {\bf Base} \\ 
                \hline 
                \multicolumn{1}{|l|}{\multirow{2}{*}{\bf General}}  & Unified & 58.30           & 59.43(M)  & 61.46             & 63.48(M)      & \textbf{59.80}    & 57.02(M)\\
                \cline{2-8}
                \multicolumn{1}{|l|}{}                              & Divided & 58.65           & 61.27(M)  & 61.26             & 63.80(M)      & \textbf{62.20}    & 50.66(M)\\
                \hline
                \multicolumn{1}{|l|}{\multirow{2}{*}{\bf Field}}    & Unified & 69.08  & 69.00(M)  & 74.72             & 76.48(M)      & 63.63             & 64.49(M)\\ 
                \cline{2-8}
                \multicolumn{1}{|l|}{}                              & Divided & 72.63           & 73.72(M)  & 77.49             & 77.74(M)      & 50.91             & 53.20(m)\\
                \hline
                \multicolumn{1}{|l|}{\multirow{2}{*}{\bf Oversight}}& Unified & 72.28           & 73.00(M)  & 66.03             & 68.10(M)      & 73.61             & 73.75(M)\\ 
                \cline{2-8}
                \multicolumn{1}{|l|}{}                              & Divided & 67.72           & 68.30(M)  & \textbf{69.30}    & 67.98(M)      & 70.60             & 71.13(M)\\ 
                \hline
                \multicolumn{1}{|l|}{\multirow{2}{*}{\bf Authorization}}& Unified & 55.72 & 55.63(M)& 52.01            & 57.05(M)      & \textbf{56.62}    & 55.40(M)\\ 
                \cline{2-8}
                \multicolumn{1}{|l|}{}                              & Divided & 59.25           & 59.96(M)  & 59.70             & 62.20(M)      & 59.30    & 59.10(M)\\ 
                \hline
                \multicolumn{1}{|l|}{\multirow{2}{*}{\bf Nomination}}& Unified  & \textbf{62.55}& 50.52(m)  & 62.43             & 66.29(M)      & \textbf{62.08}    & 54.38(m)\\ 
                \cline{2-8}
                \multicolumn{1}{|l|}{}                              & Divided & 57.00  & 55.87(m)  & 59.75             & 60.75(M)      & 100.00             & 100.00(m)\\ 
                \hline
            \end{tabular}
            \caption{Accuracy on predicting party standing (majority vs. minority) from question utterances across hearing types, comparing unified and divided governments. The Base column reports the majority-class in the test set, and are annotated either (M) or (m) to indicate that a majority of the utterances belonged to speakers from the chamber’s majority or minority party respectively. Results are grouped based on Democrat and Republican presidencies, while the `All' column represents all governments.}
            \label{tab:UniDivGovMajorityResultsHT}
        \end{table*}

    \begin{table*}[p]
        \centering
        \footnotesize
        \begin{tabular}{|c|c|c|c|c|c|c|c|l|}
            \hline
            \textbf{Llama}  & \textbf{Gemma} & \textbf{Deepseek}    & \textbf{Qwen} & \textbf{Mistral}  & \textbf{LLM Baseline} & \textbf{Hearing Type} \\\hline
            \textbf{55.89}  & \textbf{58.60} & \textbf{55.62}       & 50.31         & \textbf{54.15}    & 51.89(R)      & General\\ \hline
            50.85           & 58.05          & 44.58                & 40.12         & 53.53             & 59.79(R)      & Field\\ \hline 
            58.82           & \textbf{66.85} & 56.94                & 45.37         & \textbf{61.64}    & 61.51(R)      & Oversight\\ \hline
            \textbf{53.15}  & \textbf{62.16} & \textbf{55.02}       & 46.58         & \textbf{51.82}    & 51.46(R)      & Authorization\\ \hline
            \textbf{64.29}  & 44.12          & 54.05                & 49.53         & 50.00             & 59.32(D)      & Nomination\\ \hline
              0.86          &  0.55          &  4.91                &  3.58         & 2.58 & - & Not Answered by LLM\\\hline
        \end{tabular}
        \caption{Zero-shot predictions of questioner party affiliation from question utterances (with the speaker's name removed), across hearing types, using various LLMs. The Baseline column reports the majority-class baseline.}
        \label{tab:Party_LLM_Committees}
    \end{table*}

        \begin{table*}[p]
            \centering
            \footnotesize
            \setlength\tabcolsep{3pt}
            \begin{tabular}{|l|l|cccccc|}
                \hline
                \textbf{\bf Hearing Type}                           & Government      & \textbf{Llama}  & \textbf{Gemma} & \textbf{Deepseek} & \textbf{Qwen} & \textbf{Mistral} & \textbf{Base}\\ 
                \hline 
                \multicolumn{1}{|l|}{\multirow{2}{*}{\bf General}}  & Unified         & \textbf{55.08} & \textbf{53.17}  &  \textbf{54.50}  & 49.70         & \textbf{53.70}   & 50.16(D)\\     
                \cline{2-8}
                \multicolumn{1}{|l|}{}                              & Divided         & \textbf{57.46} & \textbf{57.65}  &  \textbf{57.50}   & 51.13         & \textbf{56.77}   & 52.48(R)\\      
                \hline
                \multicolumn{1}{|l|}{\multirow{2}{*}{\bf Field}}    & Unified         & 46.30          & 38.89           & 45.10   & 32.08         & \textbf{0.00}    & 58.33(R)\\     
                \cline{2-8}
                \multicolumn{1}{|l|}{}                              & Divided         & 54.12          & \textbf{59.14}  & 48.72             & 45.68         & \textbf{66.67}   & 57.30(R)\\ 
                \hline
                \multicolumn{1}{|l|}{\multirow{2}{*}{\bf Oversight}}& Unified         & \textbf{61.90} & 56.19           & 55.56             & 46.15         & \textbf{66.67}   & 60.18(R)\\ 
                \cline{2-8}
                \multicolumn{1}{|l|}{}                              & Divided         & \textbf{69.44} & 50.00           & 54.29            & 52.94         & \textbf{100.00}  & 58.54(D)\\ 
                \hline
                \multicolumn{1}{|l|}{\multirow{2}{*}{\bf Authorization}}& Unified     & \textbf{55.70} & 45.00           & 46.05             & 42.86         & \textbf{80.00}   & 50.59(D)\\ 
                \cline{2-8}
                \multicolumn{1}{|l|}{}                              & Divided         & \textbf{53.78} & 50.41           & \textbf{66.17}   & 50.83         & 40.00            & 51.54(R)\\ 
                \hline
                \multicolumn{1}{|l|}{\multirow{2}{*}{\bf Nomination}}& Unified        & \textbf{63.64} & 48.48           &  56.06    & 50.00         & 50.00            & 61.19(D)\\ 
                \cline{2-8}
                \multicolumn{1}{|l|}{}                              & Divided         & \textbf{61.76} & 50.00           &  50.00           & 42.42         & 50.00            & 56.41(D)\\ 
                \hline
            \end{tabular}
            \caption{Predictions of party affiliation, Democrat (D) vs. Republican (R) from question utterances across hearing types, comparing unified and divided Congress settings. The Base column reports the majority-class baseline and an annotation of (R) denotes that speakers from the Republican party have the majority number of utterances, (D) denotes Democratic speakers hold the majority class in the dataset.}
            \label{tab:UniDivGovPartyResultsHT}
        \end{table*}

\end{document}